\documentclass[12pt]{article}
\pdfoutput=1 
\usepackage[margin=0.95 in]{geometry}
\usepackage{bm}
\usepackage{amsthm}
\usepackage{breqn}
\usepackage{amssymb,amsfonts}
\usepackage[all]{xy}
\usepackage{graphicx}
\usepackage{amsmath}
\usepackage{mathtools}
\usepackage{float}
\usepackage{array}
\usepackage{mathtools}
\usepackage{mathrsfs} 
\usepackage{comment}
\usepackage{hyperref}
\usepackage{cite}
\usepackage[multiple]{footmisc}
\usepackage{physics}
\usepackage{xfrac}
\usepackage{caption}
\usepackage{slashed}
\usepackage{dsfont}
\usepackage[all]{xy}
\usepackage{graphicx}
\usepackage{float}
\usepackage{array}
\usepackage{tikz}
\usetikzlibrary{cd}
\usepackage{tikz-3dplot}
\numberwithin{equation}{section}
\numberwithin{table}{section}

\newcommand{\curlyA}{\mathcal{A}}

\newcommand{\curlyC}{\mathcal{C}}

\newcommand{\curlyL}{\mathcal{L}}
\newcommand{\curlyN}{\mathcal{N}}

\newcommand{\A}{\mathsf{A}}

\newcommand{\NN}{{\mathbb N}}

\begin{document}

\hypersetup{pageanchor=false}
\begin{titlepage}
\vbox{\halign{#\hfil    \cr}}  
\vspace*{15mm}
\begin{center}
{\Large \bf The String Dual to Two-dimensional Yang-Mills Theory}

\vspace*{3mm}

{\Large \bf Revisited}
\vspace*{10mm} 

{\large Lior Benizri and
Jan Troost}
\vspace*{8mm}

Laboratoire de Physique de l'\'Ecole Normale Sup\'erieure \\ 
 \hskip -.05cm
 CNRS, ENS, Universit\'e PSL,  Sorbonne Universit\'e, \\
 Universit\'e  Paris Cit\'e 
 \hskip -.05cm F-75005 Paris, France	 

\vspace*{0.8cm}
\end{center}

\begin{abstract}{We propose that chiral two-dimensional Yang-Mills theory on a Riemann surface $X$ is dual to a deformed stationary subsector of the Gromov-Witten theory of $X$.  
Firstly, we argue that the algebraic structure that underlies the large $N$ limit of the chiral gauge theory is a partial permutation Frobenius algebra of observables which codes covering maps of all degrees simultaneously. 
Secondly, we exploit the Gromov-Witten/Hurwitz correspondence to interpret chiral Yang-Mills theory as a finite deformation of a Gromov-Witten theory by an area-dependent transposition interaction and an operator that codes a compactification of Hurwitz space. The proposed string dual manifestly includes an integral over the moduli space of Riemann surfaces as well as the identification of closed string states as completed cycles. 
}  
\end{abstract}

\end{titlepage}

\hypersetup{pageanchor=true}

\setcounter{tocdepth}{2}
\tableofcontents

\section{Introduction} 
Two-dimensional Yang-Mills theory provides an interesting case study for the proposed relation between large $N$ gauge theory and string theory
\cite{tHooft:1973alw,Gross:1992tu,Gross:1993hu,Gross:1993yt,Cordes:1994fc}. For a gauge theory with gauge group $U(N)$, one can rearrange the Feynman diagram expansion of observables in an expansion in terms of the genus of a Riemann surface that one would like to interpret as the world-sheet of a string \cite{tHooft:1973alw}. Formulating the corresponding string theory from first principles is a challenge. Topological string theory proposals exist for the dual of two-dimensional Yang-Mills theory \cite{Horava:1993aq,Cordes:1994sd}. See also \cite{Aharony:2023tam} for a recent contribution on how to describe the dual string theory in more detail.

In this paper, we wish to increase the conceptual clarity of the correspondence between two-dimensional Yang-Mills theory and a topological string theory. The main point will be to provide a dual string theory which is of a canonical form, namely a deformed  Gromov-Witten theory. The latter has a standard definition that includes an integral over the moduli space of Riemann surfaces. The manner in which we attain the string theory dual is through a known rewriting of the gauge theory, combined with a relation between a second quantized symmetric group gauge theory and Gromov-Witten theory \cite{OP1,Benizri:2024mpx}. One essential ingredient which distinguishes our approach from the literature is the treatment of the $\Omega$ points and of the transpositions. They are both viewed as finite deformations of the string theory which respectively modify the moduli space and introduce extra string interactions.  

The plan of the paper is as follows. In section \ref{Hurwitz}, we review the necessary background on the Hurwitz topological quantum field theory of $S_d$ principal bundles on Riemann surfaces. We review its grand canonical version that treats all degrees $d$ of covering maps simultaneously and how the grand canonical theory corresponds to a Gromov-Witten theory \cite{OP1,Benizri:2024mpx}. In section \ref{YMtoHurwitz}, we review the rewriting of the two-dimensional Yang-Mills partition functions in terms of a sum of contributions of degree $d$ covering maps described by Hurwitz theories \cite{Gross:1992tu,Gross:1993hu,Gross:1993yt,Cordes:1994fc}. We argue that boundary holonomies play a role similar to local observables, much like in topological field theories. We provide an explicit map between the Hurwitz observables and those of two-dimensional Yang-Mills theory. We think of
the $\Omega$ operator as well as the transposition operator as deformations in the grand canonical Hurwitz theory. 

In section \ref{ChiralYMasaStringTheory}, we clarify the stringy nature of the chiral Yang-Mills theory and prove that the treatment of $\Omega$ operators and transposition operators as deformations is consistent with its string theoretic structure.
In section \ref{DeformedGW}, we argue that the algebra underlying the chiral Yang-Mills theory is an algebra of partial permutations. We exploit the 
correspondence between Hurwitz theory and the stationary sector of Gromov-Witten theory. This enables us  to express Yang-Mills correlators as observables of a string theory, which is a Gromov-Witten theory deformed by a number of operators that describe a modification of the moduli space as well as extra string interactions. 
We conclude in section \ref{Conclusions} with a brief description of our findings, connections with the $AdS_3$ literature, and further open questions. 

Appendix \ref{Useful Identities} is a collection of useful identities for symmetric polynomials. Appendix \ref{Solving the Dictionary} contains an explicit proof of the dictionary of observables we provide in section \ref{DeformedGW}. Finally, appendix \ref{PhaseTransitions} is an application of our perspective in that it naturally suggests a phase diagram for the deformed string theory that generalizes the stringy phase transition of \cite{Douglas:1993iia,Taylor:1994zm,Crescimanno:1994eg} to a case with two extra parameters.

\section{The Hurwitz Topological Quantum Field Theory}
\label{Hurwitz}
In this section, we review the Hurwitz topological field theory. It is a two-dimensional Dijkgraaf-Witten theory \cite{Dijkgraaf:1989pz} with gauge group $S_d$. 
The grand canonical version of this theory is dual to a string
theory\cite{OP1,Benizri:2024mpx}. We describe the string theory concretely as the stationary subsector of a matter model coupled to gravity. We also discuss the relation between the relative Gromov-Witten dual and the introduction of boundaries. 
\subsection{The Hurwitz Theory at Fixed Degree}
The Hurwitz theory is the topological quantum field theory in two dimensions with gauge group equal to the symmetric group $S_d$.\footnote{Alternatively, it is the topological symmetric orbifold of a trivial theory, which contains only the identity operator.} The operators of the Hurwitz theory are in one-to-one correspondence with the conjugacy classes of the symmetric group $S_d$, or equivalently, with the partitions of $d$. The algebra of operators is the algebra $\curlyC_d$ of conjugacy classes of the symmetric group $S_d$, which is the center of the group algebra $\mathbb{C}[S_d]$. There is a basis of operators $C_{\xi}$ which is the sum of permutation elements in a given conjugacy class $\xi$. One can also think of operator insertions as adding boundaries to the surface. A given operator then corresponds to a particular holonomy condition on this boundary up to conjugacy. To study the correlators of Hurwitz theory, it is useful to introduce a Fourier transformation on the space of conjugacy classes of the symmetric group. The Fourier space is the space of shifted-symmetric functions, and the Fourier map sends a conjugacy class sum $C_{\xi}$ to the shifted-symmetric polynomial 
\begin{equation}
   f_{\xi}(\lambda):=\lvert C_{\xi} \rvert\cdot  \frac{\chi_{\lambda}(\xi)}{d_\lambda} \, ,
   \label{GeneralizedCharacters}
\end{equation}
where $\chi_{\lambda}(\xi)$ is the character of an element in the conjugacy class $\xi$ in the representation $\lambda$ of the symmetric group $S_d$, $d_{\lambda}$ is the dimension of this representation and $\lvert C_{\xi} \rvert$ is the number of elements in the conjugacy class. 
The correlators of the theory on a Riemann surface $M_G$ of genus $G$ are given by Hurwitz numbers and can be written, up to an overall normalization by the partition function $Z[M_{G}]$ \cite{Benizri:2024mpx}, as:
\begin{align}
Z[M_{G},\xi_i] &= H_{G,d}(C_{\xi_1},\dots, C_{\xi_k}) \label{OnePointBasic}\\
			   &=\sum_{\lambda \vdash d}\left(\frac{d_{\lambda}}{d!}\right)^{2-2G}
				\prod_{i=1}^k f_{\xi_i}(\lambda)\,.
\label{PartitionFunctions}
\end{align}
These observables can be interpreted as a sum on the Fourier space labeled by representations $\lambda$ with a generalized Plancherel measure. Furthermore, the operator algebra can be exploited to reduce correlation functions to the calculation of structure constants and one-point functions. In 
principle, the one-point function should be normalized by the partition function 
of the manifold in the absence of insertions. However, to simplify the comparison
between the various theories, we will ignore the issue of normalizations and refer 
to 
\begin{equation}
    \langle C_{\xi} \rangle_{G,d} := Z[M_{G},\xi]
\end{equation}
as the one-point function of the theory.
\subsection{The Extended Hurwitz Theory}
\label{ExtendedHurwitz}
In the context of the string description of two-dimensional Yang-Mills theory, one 
needs to sum over covering maps of all degrees $d$ \cite{Gross:1992tu,Gross:1993hu}.
This motivates the introduction of an overarching grand canonical version of the 
Hurwitz theories with gauge groups $S_d$ that takes into account coverings of all 
degrees simultaneously \cite{IvanovKerov,OP1,Benizri:2024mpx}.\footnote{This is a first essential supplementary tool that will distinguish our treatment of two-dimensional Yang-Mills theory from those in the literature.}
In the process, we move from a two-dimensional Dijkgraaf-Witten theory with a finite gauge group to a topological quantum field theory defined by a Frobenius algebra only.
To treat all degrees $d$ simultaneously, we introduce an operator algebra which is the algebra of orbits of partial permutations $\curlyA_{\infty}$ of an infinite set
\cite{IvanovKerov,Benizri:2024mpx}. We provide a brief review of partial permutations 
here and refer the reader to \cite{IvanovKerov,Benizri:2024mpx} for more details. 

A partial permutation is a pair $(d,\rho)$, where $d$ is a subset of $\NN$ and $\rho$ is a permutation of elements in $d$. We refer to $d$ as the support of the partial permutation.\footnote{For a given cover, the set $d$ labels the sheets of the cover. The permutation codes a reshuffling of sheets due to ramification. Some sheets may not be shuffled.} An element $\sigma$ of the symmetric group $S_{\infty}$ acts on partial permutations as
\begin{equation}
    \sigma\cdot (d,\rho)=(\sigma(d), \sigma \rho \sigma^{-1})\, .
\end{equation}
The orbits $A_r$ of the algebra of partial permutations $\curlyA_{\infty}$ under this action are of the form
\begin{equation}
    A_{r}=\sum_{(d,\rho)\in [r]} (d,\rho)\, ,
\end{equation}
where $[r]$ denotes an orbit of partial permutations that is indexed by partitions of integers. The product of partial permutations is defined by concatenating their supports  and multiplying their permutations
\begin{equation}
    (d_1,\rho_1)\cdot (d_2, \rho_2) = (d_1\cup d_2, \rho_1 \rho_2)\, .
\end{equation}
The product of orbits is defined accordingly. Once the algebra is fixed, we only further need to define the one-point functions on the algebra of partial permutations $\curlyA_{\infty}$ to fix all correlation functions. To that end, we first introduce a fixed-degree one-point function on a Riemann surface of genus $G$:
\begin{equation}
     \langle A_r \rangle_{d,G} := \binom{m_{1}(\rho)}{m_{1}(r)} \langle C_{\rho} \rangle_{d,G}\, ,
     \label{FixedDegreeOnePoint}
\end{equation}
where $\rho$ denotes the permutation obtained by trivially completing the partial permutation $r$ and $m_{1}(r)$ (respectively $ m_{1}(\rho)$) is the number of inert entries of the partial permutation (respectively permutation).
%
Note that $m_{1}(\rho)-m_{1}(r)=d-\lvert d_r \rvert$, where $d_r$ is the support of a partial permutation in the orbit $A_r$. The definition \eqref{FixedDegreeOnePoint} is 
valid for $\lvert d_r \rvert < d$. For $\lvert d_r \rvert>d$, we set $\langle A_r \rangle_{d,G}=0$. 
We now define the physical one-point function as
\begin{equation}
    \langle A_r \rangle_G = \sum_d p^d  \langle A_r \rangle_{d,G}\, .
    \label{ppOnePoint}
\end{equation}
Importantly, our grand canonical formulation of a Hurwitz theory allows
for the introduction of a chemical potential $p$ which explicitly keeps track of the degree of the maps that contribute to a given correlator. The sum over degrees can be traded for a sum over the genus of the covering surface by using the Riemann-Hurwitz formula. Let us study this for the example of partial permutations $r$ that consist of a single cycle $k$. The correlation functions of the grand canonical symmetric orbifold theory on a Riemann surface of genus $G$ are then shown to exhibit the structure of string correlators 
\begin{equation}
    \langle A_{k_1}\cdots A_{k_n} \rangle_G  = \sum_g g_s^{2g-2} \langle g_s^{-k_1+1} A_{k_1}\cdots g_s^{-k_n+1}  A_{k_n} \rangle_{g,G}\,
\end{equation}
provided that we make the identification $p=g_s^{2G-2}$ and that we define normalized partial permutations $A'_r:=g_s^{\lvert d_r \rvert} A_r$.\footnote{Note that on the torus, the genus of the Riemann surface is determined solely by the insertions. The identification $p=g_s^{2G-2}$ yields a trivial chemical potential $p=1$ and the normalization of partial permutations must be put in by hand.} This observation is consistent with the behavior of a general symmetric orbifold theory on the sphere \cite{Aharony:2024fid}.
With this normalization, the generic correlation function becomes
\begin{equation}
    \langle A'_{r_1}\cdots A'_{r_n} \rangle  = \sum_g g_s^{2g-2} \langle g_s^{\ell(r_1)} A_{r_1}\cdots g_s^{\ell(r_n)} A_{r_n} \rangle_{g}\, ,
\end{equation}
where the length $l(r)$ of a partial permutation $r$ counts the number of parts in the partition $[r]$ of $|d_r|$.

\subsection{The String Dual}
\label{StringDual}
In this subsection, we discuss the string dual to the extended Hurwitz theory \cite{OP1}. See also \cite{Benizri:2024mpx}. Firstly, we describe the string theory, which is a theory of two-dimensional gravity coupled to two-dimensional matter. 
The matter theory is a $\curlyN=(2,2)$ supersymmetric  sigma model with world-sheet $\Sigma$ and a two-dimensional, closed target space $X$ of genus $G$. The $A$-twist of the sigma model is a topological field theory whose operators are in one-to-one correspondence with the de Rham cohomology of the target space. We denote the observable associated with the volume form on $X$ by $\omega$. The topological matter model was fully solved in \cite{Witten:1989ig}. Consider next the topological string theory obtained by coupling the  model on $X$ to topological gravity \cite{Witten:1989ig,Dijkgraaf:1990qw}. Its observables are given by the gravitational descendants of the observables of the $X$ model. We will focus on the stationary sector, namely on the half infinite tower of descendants of the volume form. We denote the corresponding operators $\tau_k$ where $k \in \mathbb{N}$ labels the gravitational descendant. 
Their correlation functions compute Gromov-Witten invariants, which are defined by integration of classes on the moduli space of Riemann surfaces of genus $g$ with a 
number of marked points equal to the number of operator insertions \cite{Witten:1989ig, OP1,Hori:2003ic}.
We stress that the observables $\tau_k$ may  be thought of as vertex operators. Indeed, they are defined in terms of the pullback of the volume form on the target space to the covering space, which is identified with the world-sheet of the theory.\footnote{We note that our identification of the world-sheet of the theory with the covering space is in accord with the description of Hurwitz theory as a symmetric orbifold and the identification of the world-sheet of the string in that context\cite{Pakman:2009zz}.} 
In string theory, every vertex operator comes with a factor of the string coupling $g_s$. Therefore, we can identify vertex operators in this theory as
\begin{equation}
    V_{k} = g_s \tau_k\, .
\end{equation}
An insertion of an operator $V_k$ on a world-sheet generates a single string state.
The correlators of the string theory may then be written as
\begin{equation}
    \langle V_{k_1} \cdots V_{k_n}\rangle = \sum_{g} g_s^{2g-2} \langle V_{k_1}\cdots V_{k_n}\rangle_{g}\, ,
    \label{VertexOp}
\end{equation}
where the sum is over the genus $g$ of the world-sheet.

A backbone to our approach is that these Gromov-Witten invariants are equal to fixed-degree correlation functions of completed cycles in Hurwitz theory \cite{OP1}. Completed single cycles are linear combinations of orbits of partial permutations defined in \cite{OP1}. They form a multiplicative basis of the algebra. We write
\begin{equation}
    \bar{A}_{k}=\sum_{r} \rho_{k,r} A_{r}\, ,
    \label{CompletedCyclesDefinition}
\end{equation}
where the sum is over partitions $r$ of varying degree.\footnote{We use the characters $k,l,m,n$ of the latin alphabet to denote integers and the characters $r,s,t$ to denote partial permutations. Permutations are denoted by Greek letters.
%
}
The coefficients $\rho_{k,r}$ vanish when $\lvert d_r \rvert>k$ and yield a Kronecker delta for $\lvert d_r \rvert = k$. 
The operator dictionary in fixed-degree correlation functions between the extended Hurwitz theory and its string dual reads \cite{OP1}
\begin{equation}
    \frac{\bar{A}_{k+1}}{k!} \leftrightarrow \tau_{k} \, .
    \label{Dictionary}
\end{equation}
Plugging this duality into the string correlators \eqref{VertexOp}, we translate an integral over moduli space to a topological quantum field theory correlator:
\begin{equation}
     \langle V_{k_1} \cdots V_{k_n}\rangle =  \sum_{r_1,\dots,r_n} \prod_i \Big(g_s^{k_i+1-\lvert d_{r_i}\rvert} \frac{\rho_{k_i+1,r_i}}{k_i!} \Big) \langle  A'_{r_1}\cdots A'_{r_n}\rangle\, .
     \label{HurwitzDualityMap}
\end{equation}
Note that the factors of the string coupling $g_s$ that appear in equation \eqref{HurwitzDualityMap} are overall factors weighting different correlators. Therefore, they do not affect the genus expansion structure 
of a given correlator. Furthermore, the completed cycles appear as subdominant corrections to the  lore according to which single cycles in a symmetric orbifold theory are dual to single string states \cite{Benizri:2024mpx}. 

Conversely, since the completed cycles form a basis of the algebra of partial permutations, equation \eqref{Dictionary} can be inverted to express the correlators of simple partial permutation in terms of vertex operators expectation values. 
We first define the coefficients $\zeta_{r,s}$ as the coefficients of the partial permutations in the basis of completed cycles
\begin{equation}
    A_r = \sum_s \frac{\zeta_{r,s}}{\prod_{i=1}^{\ell(s)} (s_i-1) !} \bar{A}_s\, .
\end{equation}
We then write the substitution rule
\begin{equation}
    A'_{r} \leftrightarrow \sum_{\substack{s \,:\\ \lvert d_s \rvert\leq \lvert d_r \rvert}} g_s^{\lvert d_r \rvert-\lvert d_s \rvert} \zeta_{r,s}\,  V_{s-1}\, ,
    \label{SubRule}
\end{equation}
%
where the $V_{s-1}:=\prod_{i=1}^{\ell(s)}V_{s_i-1}$ are products of single string vertex operators.
Correlation functions involving such multi-vertex operators $V_s$ are evaluated on the moduli space that exhibits as many marked points as there are single insertions $V_{s_i}$.
%
Physically, such multi-cycle vertex operators are identified with multi-string states. Note that completed cycles also 
include single cycle corrections \cite{OP1}
which should survive in the connected correlators. Therefore, the corrections from completed cycles do not trivialize when we consider connected correlators.

We stress that the identification of the extended Hurwitz correlators with the Gromov-Witten correlation functions \cite{OP1} is an important non-trivial step in the process of identifying a string dual to two-dimensional Yang-Mills theory. It identifies an ordinary topological field theory correlator with a well-defined integral over the moduli space of Riemann surfaces. That these highly different concepts coincide is proven through equivariant localization \cite{OP1,Okounkov:2002cz}. Exploiting this non-trivial correspondence is the second new tool that distinguishes our approach from the literature. 

To close this section, we explain how boundaries with prescribed boundary conditions can be incorporated in the gauge/string duality. Consider a target space $X$ with $m$ marked points. Hitherto, we have thought of marked points as sites of bulk operator insertions.  The operator insertions in the gauge theory translate to vertex operators in the string theory. The requirement that a basis of bulk string vertex operators be multiplicative then imposes that single cycles be completed. However, in this simple topological setting, these operator insertions can equivalently be interpreted as boundaries (with an associated holonomy). 
The  marked points then correspond to boundary conditions\footnote{This simple string theory behaves as a finite-dimensional theory and boundary states for the closed string match the vertex operator insertions.} and there is no need to complete the cycles. In other words, boundary conditions in a string theory should be treated as standard Hurwitz insertions on marked points. In Gromov-Witten theory, these insertions (in contrast to the bulk gravitational descendants of cohomology classes) are described as relative Gromov-Witten theory. 
The relative theory computes Gromov-Witten invariants on surfaces with further specified ramification profiles $\eta_i$ in $m$ marked points -- see e.g. \cite{OP1} --:
\begin{equation}
    \langle \prod_{i=1}^{s} \tau_{k_i}, \eta_1, \dots, \eta_m \rangle_{g,d}\, .
\end{equation}
To summarize, we have described two dual points of view: boundary conditions for open strings specified by ordinary partial permutations  and  closed string states corresponding to completed cycles.

Relative Gromov-Witten theory is particularly useful to study Wilson lines. As we shall review in section \ref{WilsonLoops} in the context of Yang-Mills theory, Wilson lines are computed by gluing manifolds with boundaries and specifying a character of the gauge group on that boundary. It will be more natural to think of this procedure in terms of boundaries rather than bulk operator insertions.

\section{From Two-dimensional Yang-Mills to Hurwitz 
}
\label{YMtoHurwitz}

In this section, we  explain the link between $U(N)$ Yang-Mills and Hurwitz theory in detail. We review how the correlators of two-dimensional Yang-Mills theory can be written in terms of symmetric group quantities \cite{Gross:1992tu,Gross:1993hu,Gross:1993yt}. We then discuss the large $N$ asymptotic expansion of the theory. Finally, we define a deformed Hurwitz theory that captures the perturbative expansion of observables in chiral Yang-Mills theory. The technical rewriting of correlators is largely taken from \cite{Gross:1992tu,Gross:1993hu,Gross:1993yt,Cordes:1994fc} and \cite{Novak:2023bft}. We carefully discuss the role of the degree of the covering maps, the necessity to take the large $N$ limit first as well as a natural interpretation of the operator $\Omega$ as a deformation of the theory. 

\subsection{The Observables of Two-dimensional Yang-Mills Theory}
\label{YMObservables}
Two-dimensional Yang-Mills theory is invariant under area-preserving diffeomorphisms \cite{Witten:1991we} and therefore has  no local observables. It is a prototypical example of an area dependent quantum field theory \cite{Runkel:2018uls}, which is a mild generalization of a topological quantum field theory. 
The observables of two-dimensional Yang-Mills theory are Wilson loops on the one hand and on the other hand partition functions on Riemann surfaces with boundaries with prescribed holonomies. From the point of view of topological field theories, the former corresponds to line defects and the latter to insertions of local operators. Indeed, the boundary holonomies are indexed by conjugacy classes of the unitary group. In a topological setting, one can proceed to shrink the boundaries down to singular points characterized by corresponding insertions of conjugacy classes of the unitary group. The situation is akin to the treatment of closed string vertex operators on the world-sheet in string theory. We will adopt this same point of view here, and treat the partition functions of Yang-Mills with boundaries as correlators of local insertions.\footnote{In particular, one can define a hole operator algebra. Holonomies $U_1$ and $U_2$ may be replaced in all correlation functions by linear combinations of a single holonomy $U_3$. This perspective is developed further at the end of section \ref{TheDualityMap}.} In this paper, we focus on this class of observables. We remark on Wilson loops in section \ref{WilsonLoops}.

The correlators of  pure Yang-Mills theory with gauge group $U(N)$ and coupling constant $g_{YM}$ on a Riemann surface of genus $G$ and area $A$ read -- see e.g. \cite{Blau:1991mp} for a pedagogical derivation -- : 
\begin{equation}
	Z[M_G,U_i]=\sum_{\lambda\in Y_N} \left(\prod_{i=1}^{k} s_{\lambda}(U_i)\right) (\dim W_N^{\lambda})^{2-2G-k} \exp\left(-\frac{g_{YM}^2A}{2N}\, c_{2}(\lambda)\right)\, ,
    \label{YMPartitions}
\end{equation}
where the sum runs over representations $W_N^{\lambda}$ of the unitary group which are indexed by Young diagrams $\lambda$ with at most $N$ rows.
The set of such diagrams is denoted $Y_N$. Moreover, the partition functions are functions of the holonomy classes $U_i$ of the gauge field on the boundaries. We denoted by $s_{\lambda}(U_i)$  the character of an element of the conjugacy class $U_i$ in the representation $W^{\lambda}_N$. Finally, $c_{2}(\lambda)$ is the quadratic Casimir of the representation  $W^{\lambda}_N$. Note that the coupling constant $g_{YM}$ and the area $A$ always appear in the correlators (and in the Wilson lines) in the  combination $g_{YM}^2A$. 
In other words, the dimensionless coupling constant of the theory is $g_{YM}^2A$.

It will be useful to split the sum over representations into two parts and write
\begin{equation}
    \sum_{\lambda\in Y_N} = \sum_d \sum_{\lambda\in Y^d_N}\, ,
\end{equation}
where $Y^d_N$ is the set of Young diagrams of less than $N$ rows and with $d$ 
boxes. We refer to the quantities obtained by only carrying out the second sum, 
i.e. by restricting the sum in equation \eqref{YMPartitions} to Young diagrams with exactly $d$ boxes, as fixed-degree correlators. We will denote these correlators $Z_d$. 

Let us argue that fixed-degree correlators are sensible in the whole theory -- they manifestly suffice to compute the whole theory. Indeed, the quantities $Z_d$ are observables. To prove this, we can take a dual perspective on boundaries and associate to each boundary of the manifold an irreducible representation of the gauge group $U(N)$ -- see e.g. \cite{Witten:1991we}. 
The representations must coincide to ensure that resulting correlators do not vanish \cite{Witten:1991we}. We index the unique irreducible representation 
at hand by a Young diagram $\mu$ and write
\begin{align}
    Z_{\mu}[M_{g,k}] &=\int dU_1\dots dU_k\cdot Z[U_1,\dots, U_k]\cdot s_{\mu}(U_1^{-1})\cdots  s_{\mu}(U_k^{-1})\\
    &=(\dim W_N^{\mu})^{2-2G-k} \exp\left(-\frac{g_{YM}^2A}{2N}\, c_{2}(\mu)\right)\, .
\end{align}
The fixed-degree correlators $Z_d$ are linear combinations of the partition functions $Z_{\mu}$, which implies that they are observables. This proves that these correlators are necessary and sufficient ingredients to describe the full theory.
From here on out, we refer to the set of partition functions $Z_d$ of Yang-Mills correlators at fixed-degree $d$ as the degree $d$ sector of the theory.

\subsection{The Schur-Weyl Dictionary}
Our aim in this section is to review how the Schur-Weyl duality can be leveraged to explicitly rewrite the observables of two-dimensional Yang-Mills theory in terms of symmetric group quantities\cite{Gross:1992tu,Gross:1993hu,Cordes:1994fc,Novak:2023bft}. 
Our goal is to claim that a certain sector of two-dimensional Yang-Mills theory at large $N$ is captured by a deformed Hurwitz theory.

We recall that representations of the symmetric group $S_d$ are indexed by Young diagrams with exactly $d$ boxes, and representations of the unitary group $U(N)$ relevant to the study of two-dimensional Yang-Mills theory are indexed by Young diagrams with at most $N$ rows \cite{Cordes:1994fc}. The Schur-Weyl duality relates various quantities of unitary group and symmetric group representations indexed by
the same Young diagram. 
Consider first the  equation for the quadratic Casimir of a representation of the unitary group $U(N)$ indexed by a Young diagram $\lambda$ with $d$ boxes : 
\begin{equation}
    c_{2}(\lambda)=dN+2 f_2(\lambda)\, .
    \label{Splitting}
\end{equation}
The function $f_2(\lambda)$ denotes the Fourier transform of the conjugacy class of transpositions, as defined in equation \eqref{GeneralizedCharacters}. The exponential term may then be split into two parts and the correlators can be rewritten as
\begin{equation}
	Z[M_G,U_i]=\sum_{d} q^{d} \sum_{\lambda\in Y^d_N} \frac{s_{\lambda}(U_1)\cdots s_{\lambda}(U_k)} {(\dim W_N^{\lambda})^{2g-2+k}} \exp\left(-\frac{g_{YM}^2 A}{N}\, f_{2}(\lambda)\right)\, ,
    \label{NotYetChiral}
\end{equation}
where $q:=e^{-\frac{g_{YM}^2 A}{2}}$. Next, we write the dimensions of the unitary group representation $W^{\lambda}_N$ as
\begin{equation}
	\dim W^{\lambda}_N=\frac{N^d}{d!}\Omega_{\frac{1}{N}}(\lambda) \, d_{\lambda}\, ,
    \label{dimension}
\end{equation}
where $d_{\lambda}$ is the dimension of the irreducible representation of the symmetric group indexed by $\lambda$. The function $\Omega_{s}$ is the Fourier transform of the exponential distance operator defined as \cite{Novak:2023bft} \footnote{In practice, we often think of $\Omega$ as a linear combination of conjugacy classes. Nevertheless, one can think instead of its Fourier transform $\Omega(\lambda)$ which is an operator on the set of representations of $S_d$ \cite{Novak:2023bft}. Our notation does not differentiate these two perspectives on $\Omega$.
}
\begin{align}
    \Omega_s :=&\, \sum_{\eta\vdash d} s^{\lvert \eta \rvert} C_{\eta}\\
    =&\, C_e + s C_2 + s^2 (C_3+C_{2,2})  + \mathcal{O}(s^3)\, \, ,
\end{align}
where $\lvert \eta \rvert$ denotes the charge of the permutation $\eta$, namely we have $\lvert \eta \rvert := d-l(\eta)$.  The operator $\Omega_s$ depends on a free parameter $s$.
For generic values of the parameter $s$, the exponential distance operator is invertible. Indeed, if one writes $\Omega_s=C_e+s \omega$, then 
\begin{equation}
    \Omega_s^{-1} = C_e - s \omega + s^2 \omega^2 + \dots
\end{equation}
This series can be shown to converge absolutely when $s\neq 1, \dots, d-1$ \cite{Novak:2023bft}. Conversely, if the parameter $s$ takes one of these values, then the operator $\Omega_s^{-1}$ becomes singular. The value of the operator $\Omega_{s}^{-1}$  on partitions of length smaller than $N$ is still well-defined but it diverges for partitions of length larger than $N$. This  is merely a consequence of the fact that we take  fictitious representations of $U(N)$ to have vanishing dimension, as can be seen from equation \eqref{dimension}. 
The explicit algebraic description of the operators $\Omega_s^{\pm 1}$ provides a direct hands-on description of the computational consequences of the compactification of Hurwitz spaces suggested in \cite{Cordes:1994fc}. 

To complete the Schur-Weyl dictionary, we discuss how insertions in Yang-Mills translate to the Hurwitz theory. We have the identity
\begin{equation}
    s_{\lambda}(U)=\frac{d_{\lambda}}{d!}\sum_{\alpha\in Y^d} p_{\alpha}(U) f_{\alpha}(\lambda)\, 
    \label{SchurToPower}
\end{equation}
where the power-sum polynomials $p_\alpha(U)$ are defined as 
\begin{equation}
    p_{\alpha}(U)=\prod_{i=1}^{l(\alpha)} \tr U^{\alpha_i}\, .
\end{equation}
One recognizes in equation \eqref{SchurToPower} the change of basis  that relates Schur functions and power sum polynomials. 
Substituting all of these identities in the partition function of the Yang-Mills theory we obtain
\begin{equation}
	Z[M_G,U_i]=\sum_d (q N^{2-2G-k})^d\, Z_{d}[M_G,U_i]\, ,
    \label{dSum}
\end{equation}
where we defined the fixed-degree result
\begin{multline}
     Z_{d}[M_G,U_i]:= \sum_{\lambda\in Y_N^d} \left(\frac{ d_{\lambda}}{d!}\right)^{2-2G} \prod_{i=1}^{k} \left(\sum_{\alpha\in Y^d} p_{\alpha}(U_i) f_{\alpha}(\lambda) \cdot \Omega_{1/N}^{-1}(\lambda)\right)\\ (\Omega_{1/N}(\lambda))^{2-2G} \exp\left(-\frac{g_{YM}^2 A}{N}f_{2}(\lambda)\right)\, .
     \label{ExactCorrelators}
\end{multline}
Note that this definition of $Z_d$ differs from the one introduced in section \ref{YMObservables} by a factor $q^d N^{(2-2G-k)d}$. From now on, $Z_d$ will always denote the factored out fixed-degree correlator. We stress that equation \eqref{dSum} is an exact expression for the correlators of the full two-dimensional Yang-Mills theory in terms of symmetric group quantities. 

\subsection{The Gross-Taylor Expansion}
\label{GTExpansion}
Consider  the fixed-degree correlator \eqref{ExactCorrelators}. It features a sum over Young diagrams of the Fourier transform of conjugacy classes weighted by a generalized Plancherel measure -- it seems it can be interpreted as a complicated Hurwitz correlator. Nevertheless, there is an obstacle to this interpretation, namely that the sum over Young diagrams runs only over diagrams with less than $N$ rows. If we wish to reach a symmetric group correlator interpretation furthermore to be identified with a string correlation function, it is imperative to take the large $N$ limit first. Therefore, we turn to the study of the large $N$ asymptotic expansion of Yang-Mills theory. We will explain how to perform the expansion as well as its regime of validity. We thus prepare the ground for the following sections where we will match the perturbative expansion of Yang-Mills theory with the correlators of a perturbatively defined string theory.

To produce the asymptotic expansion, we expand each $N$-dependent term inside the sum over partitions in equation \eqref{ExactCorrelators}. However, on Riemann surfaces of genus $G\leq 1$, this expansion only captures a chiral half of the full theory. Indeed, the Young diagrams with $N$ boxes or more appear to be exponentially suppressed in the expansion, because the factor $q^d$ is of order $\mathcal{O}(e^{-N g_{YM}^2 A})$ for these diagrams.\footnote{The nature of the instantonic correction is interesting. It is both similar to a world-sheet instanton correction with exponent proportional to the area, but also to a non-perturbative correction in the string coupling because the wrapping number is of order $N=1/g_s$. Using also the identification \eqref{StringTension}, the correction can be interpreted as due to a D-brane instanton.} 
Yet some of these representations are then wrongfully convicted: they are only exponentially suppressed as an artifact of our splitting of the exponential term (\ref{Splitting}). Indeed, for certain special diagrams \cite{Gross:1993hu}, the 
terms $dN$ and $2f_{2}(\lambda)$ conspire to give a result of order $N$ and thus 
an exponential term of order $1$, even though both of these terms are of order $N^2$. This can be seen explicitly by expanding the term $f_{2}(\lambda)$ \cite{Gross:1993hu}. We define the chiral theory as the naive asymptotic expansion of the full two-dimensional Yang-Mills theory.
In \cite{Gross:1993hu}, all representations with a Casimir of order $\mathcal{O}(N)$ were identified, and it was argued that (the asymptotic expansion of) the full theory can be encoded into two coupled copies of a chiral theory. In this paper, we do not discuss the full Yang-Mills theory on Riemann surfaces of genus $G\leq 1$ any further --  we shall focus instead on deriving a string dual for the chiral Yang-Mills theory. 

The situation is different for higher genus Riemann surfaces \cite{Gross:1992tu}. There, representations indexed by diagrams comprising $N$ boxes or more are exponentially suppressed regardless of whether we split the Casimir term in two. Indeed, for each value of $d$, the dominant term in the asymptotic expansion in $N$ is of order $N^{-d}$. Therefore, diagrams with $d\sim N$ boxes are exponentially suppressed as $e^{-N \log N}$. The perturbation theory of the chiral and non-chiral Yang-Mills theory therefore coincide. 
In summary, the perturbative string theory we introduce in this paper is dual to the full Yang-Mills theory on Riemann surfaces of genus $G>1$, and to a chiral half of the Yang-Mills theory for $G\leq 1$.

We  only provided a perturbative definition of the chiral Yang-Mills theory in its large $N$ limit.\footnote{A non-perturbative definition of the chiral theory can perhaps be achieved by restricting the sum over degrees $d$ in equation \eqref{NotYetChiral} to degrees bounded above by $d\leq N$.}
Accordingly, our method to derive a string dual is to look for a string theory whose correlators match this asymptotic expansion. To that effect, we attempt in the next section to interpret the right-hand side of equation \eqref{ExactCorrelators} as 
Hurwitz numbers. This requires lifting the constraint on the number of rows in the Young diagram
\begin{equation}
    \sum_{\lambda\in Y^d_N} \rightarrow \sum_{\lambda\in Y^d}\, ,
\end{equation}
which is synonymous with adding infinitely many exponentially suppressed terms. Many 
of the added diagrams are ill-defined, because the value of the exponential distance operator $\Omega^{-1}_{1/N}(\lambda)$ diverges on Young diagrams $\lambda$ with more than $N$ rows.\footnote{The only observables which are not affected by this divergence are the partition functions on the sphere and the torus, as well as the one and two point functions on the sphere.}
This divergence must be regularized. This may be done by shifting the parameter $u=1/N$ of the exponential distance operator by $\epsilon$, discarding the exponentially suppressed terms in the large $N$ limit, then taking $\epsilon$ to $0$. We implicitly adopt this prescription throughout the remainder of the paper.

Finally, if either the area or the coupling vanishes, it is no longer true 
that certain representations of the unitary group are exponentially
suppressed.\footnote{Nevertheless, our prescription to regularize the asymptotic expansion holds, at least on Riemann surfaces of genus $G>1$, because diagrams with $N$ boxes or more are exponentially suppressed by factors $N^{-N}$, as explained above.}
Therefore, in the topological limit, the chiral theory and the full theory coincide.
As a result, the topological limit of the string theory we introduce in this paper is 
dual to the full topological Yang-Mills theory in two dimensions.

\subsection{Chiral Yang-Mills Theory as a Deformed Hurwitz Theory}
\label{Premise}
We have all the necessary tools in hand to define a deformed Hurwitz theory that is manifestly dual to two-dimensional chiral Yang-Mills theory. As suggested at the end of section \ref{YMObservables}, we will focus on the degree $d$ sector of chiral Yang-Mills theory and postpone the discussion of the sum over degrees. Firstly, we consider the fixed-degree correlators of the chiral theory in the strict large $N$ limit: 
\begin{align}
    Z^{+}_{d}(U_1,\dots,U_k) &=\sum_{\alpha_1,\dots, \alpha_d\in Y^d} \left(\prod_{i=1}^{k} p_{\alpha_i}(U_i)\right)  \sum_{\lambda\in Y_d} \left(\frac{ d_{\lambda}}{d!}\right)^{2-2g} \prod_{i=1}^{k}  f_{\alpha_i}(\lambda)
    \label{StrictLargeN}\\
    &=H_d\left(\sum_{\alpha} p_{\alpha}(U_1) C_{\alpha},\dots,\sum_{\alpha} p_{\alpha}(U_k) C_{\alpha}\right)
    \, .
\end{align}
The subscript $+$ denotes the chiral correlators. The sum over Young diagrams on the right-hand side is manifestly a linear combination of Hurwitz number with coefficients that are polynomials in the holonomies. As a result, the degree $d$ sector  reduces in the strict large $N$ limit to a Hurwitz theory with gauge group $S_d$. The dictionary of observables sends a holonomy class $U$ in the Yang-Mills theory to an operator $\sum_{\alpha} p_{\alpha}(U) C_{\alpha}$ in the Hurwitz theory.

When the $1/N$ corrections are turned on, the right-hand side of equation \eqref{ExactCorrelators} features additional terms compared to the limit \eqref{StrictLargeN}: simple branch points on the one hand, and $\Omega$ operators on the other. One difficulty in establishing a correspondence between two-dimensional Yang-Mills theory and Hurwitz theory lies in the treatment of these $\Omega$ factors.

The novelty in our approach will consist in treating these factors as arising from a deformation of Hurwitz theory.\footnote{The papers \cite{Cordes:1994fc,Horava:1993aq} have treated  area terms in the Yang-Mills partition function by perturbing the conjectured dual string theory by an area operator. We modify the treatment and extend the strategy to the $\Omega$ points. Furthermore, we note that the Yang-Mills point in theory space corresponds to finite deformations by transpositions and $\Omega$ points.
}
There is a  modification of the moduli space captured by a factor $\Omega^{2-2G}$ in the partition function \cite{Cordes:1994fc}. The renormalization of the observables accounts for the additional $\Omega^{-k}$ factor that appears in the Gross-Taylor expansion. 

We proceed to define the appropriate deformations of Hurwitz theory. We first incorporate the single branch points by adding the extra  term
\begin{equation}
      \curlyL \to \curlyL - g_s t\cdot C_2 \, 
    \label{TransDeformation}
\end{equation}
to the exponent of the path integral measure formally represented by a Lagrangian $\curlyL$. The factor $g_s$ denotes the string coupling constant in the dual theory and is identified with $1/N$. It is featured in the perturbation to ensure that the deformation is small for small values of $t$. Indeed, as we shall see in section
\ref{AStringyDeformation}, whether the deformation parameters are large or small is 
determined by their value relative to the string coupling.
Furthermore, factoring out $g_s$ allows us to  interpret the parameter $t$ as the area of the target space in units of the Yang-Mills coupling $g_{YM}$. We now introduce the $\Omega$ deformation. Recall that the operator $\Omega$ has a natural expression as a linear combination of conjugacy classes. Furthermore, we have the first order estimate $\Omega_{g_s u}=1+\mathcal{O}(g_s u)$, which implies that the logarithm of the operator is well-defined perturbatively. We write
\begin{equation}
    \curlyL \to \curlyL + \log \Omega_{g_s u}^{2-2G}\, .
    \label{OmegaDeformation}
\end{equation}
We have introduced the parameter $u$ to highlight the need to implement the regularization description of the inverse exponential distance operator described in section \ref{GTExpansion}. In terms of $u$, the prescription consists in evaluating the correlators at $u=1-\epsilon$, dropping the exponentially suppressed terms from the expansion, then taking the limit $\epsilon \to 0$.

The above symbolic representations can be made precise by stating that the correlators of the deformed Hurwitz theory at order $d$ read
\begin{equation}
    \langle \prod_{i=1}^{k} C_{\alpha_i} \rangle_d^{t,u} 
    =\langle \left(\prod_{i=1}^{k} C_{\alpha_i}\right) e^{-g_s t C_2}\, \Omega_{g_s u}^{2-2G}  \rangle_d\, ,
    \label{DeformedHurwitz}
\end{equation}
where the superscript $t,u$ on the left-hand side indicates that the parameters of the deformation are $t$ and $u$, respectively. To simplify the notations, we shall  use a superscript ${\text{def}} $ to denote the Hurwitz theory deformed by the parameters $t=g_{YM}^2 A$ and $u=1$, up to a regularization procedure. Comparing with the expression for the fixed-degree Yang-Mills correlators in the large $N$ limit \eqref{ExactCorrelators}, we find that agreement between the correlators of both theories necessitates that the operators of the Hurwitz theory be renormalized as 
\begin{equation}
    C_{\alpha}\to \hat{C}_{\alpha}:=\Omega_{g_s u}^{-1} C_{\alpha}\, .
\end{equation}
The inverse exponential distance operator $\Omega_{g_s u}^{-1}$ is an element 
of the conjugacy class algebra $\curlyC_d$. Therefore, the renormalized class $\hat{C}_{\alpha}$ is a well-defined operator in this algebra.

We may finally equate the  perturbative expansion of the correlators on both sides of the chiral theory 
\begin{equation}
    Z_{d}^{+}[M_G,U_i]=\langle \prod_{i} \left(\sum_{\alpha\vdash d} p_{\alpha}(U_i) \hat{C}_{\alpha}\right) \rangle_d^{\text{def}}\, ,
\end{equation}
and conclude that the large $N$ limit of the degree $d$ sector of chiral Yang-Mills theory is dual to the deformed Hurwitz theory for deformation parameters equal to $t=g_{YM}^2 A$ and $u=1$. The dictionary of observables sends a holonomy insertion $U$ in the Yang-Mills theory to an operator $\sum_{\alpha\vdash d} p_{\alpha}(U_i) \hat{C}_{\alpha}$ in the Hurwitz theory. 

\section{The Stringy Structure of Chiral Yang-Mills Theory}
\label{ChiralYMasaStringTheory}
In this section, we develop the string theory perspective on the chiral Yang-Mills theory further. Firstly, we neglect the deformations introduced in section \ref{Premise} and explain the structure of the partition function as a double expansion over the genus of the worldsheet and the instanton degree. Secondly, we discuss the normalization and operator algebra structure of the vertex operators. Lastly, we implement the deformation and show that the associated string coupling factors which seem to break the structure of the genus expansion can in fact be re-summed to preserve this structure.
\subsection{A Double Expansion}
If we  neglect the $\Omega$ points and deformation by transpositions\footnote{Note that, formally, we have defined a deformed Hurwitz theory for arbitrary values of the parameters $t$ and $u$. Therefore, equation \eqref{dSumDictionary} may be thought of as the corresponding string theory correlation functions for $t=u=0$. We discuss this aspect further in appendix \ref{PhaseTransitions}.} the partition function of chiral Yang-Mills theory on a surface with $k$ boundaries becomes:
\begin{equation}
    Z_{G}^{+}(U_1,\dots,U_k)=\sum_d (q\, g_s^{2G-2+k})^d\, \langle \prod_{i=1}^k \left(\sum_{\alpha\vdash d} p_{\alpha}(U_i) C_{\alpha}\right) \rangle_d\, .
    \label{dSumDictionary}
\end{equation}
We wish to make the string theory structure of these correlators manifest. In string theory, two expansions must be distinguished. On the one hand, there is the perturbative expansion in the string coupling $g_s$. On the other hand, there is an expansion over world-sheet instantons, with expansion parameter $e^{-\frac{A}{2\pi \alpha'}}$, where $\frac{1}{2\pi \alpha'}$ is the string tension and $A$ is the area of the embedding of the world-sheet in the target space. In our two-dimensional context, the world-sheet covers the target space and $A$ equals the area of the target space. In order to bring out the structure of the double expansion, we reshape the expression \eqref{dSumDictionary} using the Riemann-Hurwitz formula\cite{CavalieriMiles}:
\begin{equation}
    2g-2=d(2G-2)+kd-\sum_i \ell(\alpha_i)\, ,
\end{equation}
where $g$ denotes the genus of the covering surface and $\ell(\alpha_i)$ is the length of a conjugacy class, i.e. the number of disjoint cycles of any element in this class.
We write 
\begin{equation}
    Z_{G}^{+}(U_1,\dots,U_k) =\sum_d q^d \sum_{\alpha_i\in Y^d} g_s^{2g-2} \left(\prod_{i=1}^{k} p_{\alpha_i}(U_i)\right)\left \langle g_s^{l(\alpha_1)}  C_{\alpha_1} \cdots  g_s^{l(\alpha_k)} C_{\alpha_k}\right \rangle_d\, .
    \label{StringyHurwitz}
\end{equation}
Let us analyze the content of equation \eqref{StringyHurwitz}. Firstly, we recall that, as explained in section \ref{StringDual}, the covering space of Hurwitz theory is naturally interpreted as the world-sheet of a string theory. Therefore, the factor $g_s^{2g-2}$ can be interpreted as the usual weighting factor for the different world-sheet topologies. However, unlike the standard case, there is no explicit sum over all possible topologies of the world-sheet in the above expression. This a distinctive feature of the string theory at hand. Indeed, for each correlator of a Gromov-Witten theory, there is only one contributing world-sheet at fixed instanton degree $d$. Nevertheless, an insertion in the Yang-Mills theory is represented by a sum over all possible conjugacy classes 
in the Hurwitz theory, which implies that many genera come into play. 
One can make this sum over the genus of the world-sheet manifest by introducing the constrained correlators 
\begin{equation}
    \langle \cdot \rangle_{g,d}:=\langle \cdot \rangle_{d} \cdot \delta(g=g(d))\, ,
    \label{FieldToString}
\end{equation}
where $g(d)$ in the argument of the $\delta$ function is the genus of the covering surface determined by the Riemann-Hurwitz formula. It is a standard fact in topological string theory that there exist topological constraints on correlators.
In terms of the constrained correlators, the partition function \eqref{StringyHurwitz} reads
\begin{equation}
    Z_{G}^{+}(U_i) =\sum_g g_s^{2g-2} \sum_d q^d \left \langle \left(\sum_{\alpha\vdash d} p_{\alpha}(U_1) g_s^{l(\alpha)}  C_{\alpha}\right) \cdots \left(\sum_{\alpha\vdash d} p_{\alpha}(U_k) g_s^{l(\alpha)} C_{\alpha}\right)\right \rangle_{g,d}\, . 
    \label{LooksLikeAString}
\end{equation}
This is manifestly the genus expansion of a string theory and we interpret the sum over $d$ as the sum over world-sheet instanton degrees. This allows us to identify the string tension as -- see also \cite{Gross:1993yt,Horava:1993aq} --:
\begin{equation}
    \alpha'=\frac{1}{\pi g_{YM}^2}\, . \label{StringTension}
\end{equation}
\subsection{Vertex Operators}
In string theory, correlation functions on a genus $g$ world-sheet at instanton degree $d$ scale as $g_s^{2g-2+k}$, where $k$ is the number of insertions on the world-sheet.
The genus expansion is canonically written as a sum over all world-sheet topologies weighted by a factor $g_s^{2g-2}$. The remaining factor $g_s^k$ is relegated to the normalization of vertex operators: each one comes with a single factor of the string coupling $g_s$. In Hurwitz theory, the insertion of a conjugacy class $C_{\alpha}$ in 
the target space corresponds to the insertion of $\ell(\alpha)$ marked points on the worldsheet. Therefore, each such insertion must be weighted by an additional factor $g_s^{\ell(\alpha)}$, and this is indeed what we find. The physical interpretation of this finding is straightforward. By lore, single cycles correspond to single string states, and come with a single factor $g_s$.\footnote{It was noted in \cite{Lerche:2023wkj,Benizri:2024mpx}, based on \cite{OP1}, that there are interesting corrections to this lore. We discuss this aspect further in section \ref{DeformedGW}.
} %
Accordingly, multi-cycles correspond to multi-string states and the length of a conjugacy class counts the number of strings in the state defined by this conjugacy class.

This state of affairs is complicated by the presence of the coefficients $p_{\alpha}(U)$, which also scale with $N$. Consider an element $U$ of the unitary group $U(N)$ 
that acts non-trivially on a $p$-dimensional space, with $p\ll N$. Then, 
\begin{equation}
    p_{\alpha}(U)=N^{\ell(\alpha_i)} \prod_i \left(1-\frac{\Tr_p (U\rvert_p)^{\alpha_i}}{N}\right)\, ,
\end{equation}
where $U\rvert_p$ denotes the restriction of the element $U$ to its non-trivial subspace, and $\Tr_p$ the trace on this subspace.
In other words, $p_{\alpha}(U)$ scales as $g_s^{-\ell(\alpha)}$, up to corrections in $g_s$. Thus, the dominant part of the correlator on a surface of genus $g$ scales as $g_s^{2g-2}$, as expected from a Yang-Mills partition function. The identification of the correlation functions of both theories necessitates that we insert rescaled vertex operators  on the string theory side. The dictionary of observables at order $d$ reads
\begin{equation}
    U \to g_s^{-d} \left( C_e + \mathcal{O}(g_s) \right)\, .
\end{equation}
Before we move on the implementation of the deformations, we remark that in going from a field theory to a string theory correlation function in equation \eqref{FieldToString}, we have dismantled the operator algebra structure of the observables. Indeed, in the field theory,
\begin{equation}
    \langle C_{\alpha}, C_{\beta}\rangle_d = c_{\alpha \beta}^{\ \ \gamma} \langle C_{\gamma}\rangle_d\, ,
\end{equation}
where the $c_{\alpha\beta}^{\ \ \gamma}$ are the structure constants of the conjugacy class algebra $\curlyC_d$. In the string theory, this equation no longer holds, because the Riemann-Hurwitz formula is not preserved by the product of the algebra, which implies 
that the value of the genus for which $\langle C_{\gamma} \rangle_{g,d}$ is non-zero depends on $\gamma$. 
The break-down of the algebra is crucial for dealing with the deformations, as we explain momentarily.\footnote{An interesting observation is that the product of the filtered Hurwitz theory -- see e.g. \cite{Li:2020zwo} -- does preserve the Riemann-Hurwitz formula.}
Finally, we stress that in order to simplify notations, we systematically write 
\begin{equation}
    \langle C_{\alpha} C_{\beta}\rangle_{g,d}\, ,
\end{equation}
in order to refer to the correlation function
\begin{equation}
    \langle C_{\alpha}, C_{\beta}\rangle_{g,d}\, .
\end{equation}
In this paper, expressions involving products of correlators should always be understood as the insertion of two separate observables.

\subsection{A Stringy Deformation}
\label{AStringyDeformation}
We start from the undeformed partition function \eqref{dSumDictionary} and study the implementation of the deformation into the structure of the genus expansion of the string theory. Consider first the deformation by transpositions. The exponential can be expanded inside the correlation functions to write
\begin{equation}
    Z_{G}^{+}(U_1,\dots,U_k) = \sum_d g_{s}^{(2G-2+k)d} q^d \sum_r (-1)^r \frac{(g_s t)^r}{r!} \langle C_2^r \prod_{i=1}^{k} C_{\alpha_i}\rangle_{d}
\end{equation}
The Riemann-Hurwitz equation may be leveraged to write this expression as a correlator 
in a string theory. The method consists again in introducing a redundant sum over all genera together with a constrained correlator $\langle \cdot \rangle_{g,d}$:
\begin{equation}
    Z_{G}^{+}(U_1,\dots,U_k) = \sum_g g_{s}^{(2G-2+k)d+r} \sum_d  q^d \sum_r (-1)^r \frac{t^r}{r!} \langle C_2^r \prod_{i=1}^{k} C_{\alpha_i}\rangle_{g,d}\, .
\end{equation}
The correlator is non-vanishing when the genus $g$ takes the value determined by the Riemann-Hurwitz formula 
\begin{equation}
    2g-2=(2G-2+k)d-\sum_i \ell(\alpha_i)+r\, .
\end{equation}
One then identifies the power of the string coupling $g_s$ with the genus $g$ of the covering surface, up to the appropriate normalization factor for the vertex operators. 
We write:
\begin{align}
    Z_{G}^{+}(U_1,\dots,U_k) = \sum_g g_{s}^{2g-2} \sum_d  q^d \langle e^{-tC_2} \prod_{i=1}^{k} g_s^{\ell(\alpha_i)} C_{\alpha_i}\rangle_{g,d}\, .
\end{align}
The genus expansion has absorbed the factor of the string coupling $g_s$ featured in the deformation. The deformation by transpositions thus preserves the stringy structure of the correlators. In general, a deformation by conjugacy classes $C_{\eta}$ preserves the stringy structure of the correlators if each insertion comes with a factor of the string coupling raised to the power of the charge of the conjugacy class $g_{s}^{\lvert \eta \rvert}$. For transpositions, $\lvert C_2 \rvert = 1$, and each insertion $C_2$ comes indeed with a single factor $g_s$. Importantly, this means that we interpret the power expansion of the exponential as insertions of distinct observables: $\langle C_2^r C_{\alpha} \rangle_{g,d}$ denotes the correlator
\begin{equation}
    \underbrace{\langle C_2,\dots, C_2}_{r\ \text{times}}, C_{\alpha} \rangle_{g,d}\, ,
\end{equation}
as indicated in the previous section.

We move on to study the effect of $\Omega$ deformations. We recall the definition of the exponential distance operator $\Omega$:
\begin{equation}
    \Omega_{g_s u} :=\, \sum_{\eta\vdash d} (g_s u)^{\lvert \eta \rvert} C_{\eta} \, .
\end{equation}
The criterion just stated is manifestly satisfied and we deduce that the $g_s$ factor in 
$\Omega_{g_s u}$ is absorbed by the genus expansion. Furthermore, the inverse $\Omega_{g_s u}^{-1}$
can be defined by a series expansion, in which each term $C_{\eta_1}\dots C_{\eta_s}$ is again accompanied by a factor $g_{s}^{\lvert \eta_1 \rvert+\dots+\lvert \eta_s \rvert}$. As a result, the genus expansion soaks up the factors of the string coupling contained in the operator $\Omega_{g_s u}^{-1}$ as well.

In conclusion, the observables of chiral Yang-Mills theory admit a string theoretic expansion
\begin{equation}
    Z_{G}^{+}(U_1,\dots,U_k) = \sum_g g_{s}^{2g-2} \sum_d  q^d \langle \prod_{i=1}^{k} \left(\sum_{\alpha_ik} p_{\alpha}(U) g_s^{\ell(\alpha_i)} \hat{C}_{\alpha_i}\right) \rangle_{g,d}^{\text{def}}\, ,
\end{equation}
where the correlation functions of the deformed theory are defined by
\begin{equation}
    \langle \hat{C}_{\alpha_1}\cdots \hat{C}_{\alpha_k}\rangle_{g,d}^{\text{def}} =  \langle C_{\alpha_1}\cdots C_{\alpha_k}\cdot e^{-tC_2} \Omega_{u}^{2-2G-k} \rangle_{g,d} \, .
\end{equation}
We stress that by the explicit criterion we laid out, generic deformations do not preserve the string theoretic structure of correlation functions. 
We thus view the preservation of the  structure as further new and strong evidence that chiral Yang-Mills theory in two dimensions admits a string dual. 

\section{A Deformed Gromov-Witten Theory}
\label{DeformedGW}
In this section, we propose that the dual of chiral Yang-Mills theory is a deformed matter  coupled to topological gravity model.
Firstly, we argue that the algebra of partial permutations introduced in section \ref{ExtendedHurwitz} is the natural structure that underlies the symmetric group perspective on the Yang-Mills theory. We then leverage the equivalence of the extended Hurwitz theory with a Gromov-Witten theory to propose a string dual for the chiral Yang-Mills theory. We derive the dictionary of observables and identify a completed version of power sum polynomials as the dual of vertex operators. Lastly, we study how to compute Wilson loops in the string theory.
\subsection{The Partial Permutation Algebra of Yang-Mills Theory}
We have shown that the correlators of chiral Yang-Mills theory may be written 
as a sum over Hurwitz theory correlators of all degrees with insertions specified by operators in the algebra $\curlyC_d$ of conjugacy classes of degree $d$.
This observation suggests that the appropriate structure to deal with the right-hand side of equation \eqref{dSumDictionary} is one that accommodates each degree $d$ operator algebra $\curlyC_d$. The further requirement that this algebraic structure be that of a quantum field theory, and thus be an operator algebra, necessitates that a product be defined between operators of different degrees. 
Hence, one is naturally led to introduce an algebra of partial permutations
$\curlyA_{\infty}$ to interpret equation \eqref{dSumDictionary}. The required field theory is obtained by deforming the extended field theory introduced in section \ref{ExtendedHurwitz}. 
Thus far, we have only defined deformations of theories at fixed-degree -- see equations \eqref{TransDeformation} and \eqref{OmegaDeformation}. Extending such deformations to theories that capture all covering degrees at once requires care. In particular, one needs 
to define a partial permutation that behaves as the exponential distance operator $\Omega_{u}$ does inside any fixed-degree correlator. We define the appropriate operator
\begin{equation}
    \Tilde{\Omega}_u:=\sum_{r\ \text{minimal}} u^{\lvert r \rvert} A_{r}\, ,
\end{equation}
where $\lvert r \rvert := \lvert d_r \rvert - \ell(r)$ is the charge of the partial permutation $r$, and the sum is taken over all minimal partial permutations $r$, namely the partial permutations with no inert entries. This includes the neutral element
of the partial permutation algebra $A_{\empty}$ \cite{IvanovKerov}. The partial permutation $\tilde{\Omega}_u$ is defined with no reference to a specific value of the degree $d$ and indeed satisfies 
\begin{equation}
    \langle \Tilde{\Omega}_{u} \cdot (\dots) \rangle_d =  \langle \Omega_{u} \cdot (\dots) \rangle_d\, ,
\end{equation}
for all values of the degree. 

The following deformation of the extended theory thus reproduces the partition functions \eqref{dSumDictionary} of chiral Yang-Mills theory  in the absence of boundaries:
\begin{equation}
    \curlyL \to \curlyL - g_s t\, A_2 + \log \Tilde{\Omega}_{g_s u}^{2-2G} \, ,
    \label{ppDeformation}
\end{equation}
where $A_2$ is the orbit of transpositions. As for the deformed Hurwitz theory, the observables are renormalized as $\hat{A}_r:=\tilde{\Omega}_{u}^{-1} A_r$ and the correlation functions of the theory are defined by summing over all degrees and covering surface genera. We introduce the vertex operators $\hat{\A}_{r}:=g_s^{\lvert d_r \rvert} \hat{A}_{r}$ 
and write
\begin{equation}
    \langle \hat{\A}_{r_1}\cdots \hat{\A}_{r_k}\rangle^{\text{def}} := \sum_{g} g_{s}^{2g-2} \sum_d q^d \langle g_s^{\ell(r_1)}\hat{A}_{r_1}\cdots g_s^{\ell(r_k)}\hat{A}_{r_k} \rangle_{g,d}^{\text{def}}\, ,
\end{equation}
where the deformed correlator equals $\langle\, (\, \dots)\, \cdot\,  \rangle_{g,d}^{\text{def}}:=\langle \,(\, \dots)\, \cdot\, \tilde{\Omega}_u^{2-2G\,} e^{-tA_2} \rangle_{g,d}$.

\subsection{The Duality Map}
\label{TheDualityMap}
In order to extend this picture to the correlation functions, we need to establish a dictionary between the observables of chiral Yang-Mills theory and those of the deformed extended Hurwitz theory. Given an insertion $U$, we need to identify the unique element $\sum_{s} a_s(U) \hat{\A}_s$ in the algebra of partial permutations which behaves as 
\begin{equation}
    \sum_{\alpha\vdash d} p_{\alpha}(U) g_s^{\ell(\alpha)} \hat{C}_{\alpha}
\end{equation}
inside genus-$g$, degree-$d$ correlation functions, for all values of the genus and degree. Our strategy is the following. Firstly, we isolate permutations with a given non-trivial structure. That is, we look for permutations $\sigma$ obtained by completing the minimal partial permutation $\bar{s}$ to all possible degrees. For each such permutation $\sigma$, we have the equation
\begin{equation}
    \sum_{s} g_s^{\ell(s)}  a_s(U) \langle \hat{A}_s \, (\dots) \rangle_{g,d} = g_s^{\ell(\sigma)} p_{\sigma}(U) \langle \hat{C}_{\sigma} \, (\dots) \rangle_{g,d}\, ,
    \label{DictionaryCoeffEqs}
\end{equation}
where the only partial permutations $s$ relevant to the sum are the partial permutations of the form $\bar{s}\cup 1^k$, with $0\leq k\leq d-\lvert d_{\bar{s}} \rvert$. Thus, we obtain the following equations for the coefficient $a_{\bar{s}\cup 1^k}$
\begin{equation}
    \sum_{k=0}^{d-\lvert d_{\bar{s}} \rvert} g_{s}^{\ell(\bar{s})+k} a_{\bar{s}\cup 1^k}(U) \binom{d-\lvert d_{\bar{s}} \rvert}{k} = g_{s}^{\ell(\sigma)}\, p_{\sigma}(U) \, .
\end{equation}
There is an equation for each degree $d\geq \lvert d_{\bar{s}} \rvert$, and the coefficients $a_s$ thus obey an infinite system of linear equations.
Its matrix form is lower triangular and is a version of the Pascal triangle weighted by factors of the string coupling. It can be solved recursively, as explained in appendix \ref{Solving the Dictionary}. The coefficients of the partial permutation dual to an insertion $U$ are given by
\begin{equation}
    a_s(U):= p_{\bar{s}}(U) \sum_{l=0}^{m_{1}(s)} (-1)^{m_{1}(s)-l} g_{s}^{l-m_{1}(s)} \binom{m_{1}(s)}{l}  p_{1^l}(U) \, .
    \label{DictionarySolution}
\end{equation}
We conclude that the insertion of a boundary holonomy $U$ in the chiral Yang-Mills theory translates to the insertion of the observable $\sum_s a_s(U) \hat{\A}_s$ in the deformed extended Hurwitz theory. The role of trivial sheets in the string theory is played by the power-sum polynomials in the Yang-Mills theory. Each trivial sheet corresponds to a factor $\tr U$. %

An interesting consequence of this dictionary is that in the large $N$ limit, a hole operator algebra can be defined for the Yang-Mills theory. The elements of this algebra are the holonomies in the unitary group $U(\infty)$. The product is defined such that, in any given correlation function, holonomies $U_1$ and $U_2$ may be replaced by a holonomy $U_3$. We write
\begin{equation}
    Z(U_1,U_2,\dots) = \int dU_3\,  c(U_1,U_2;U_3) Z(U_3,\dots)\, .
\end{equation}
The $c(U_1,U_2;U_3)$ are structure constants for the algebra and they can be written as:
\begin{equation}
    c(U_1,U_2;U_3):=\sum_{r,s,t} \hat{g}_{rs}^{\ \ t} g_s^{\lvert d_r \rvert+\lvert d_s \rvert-\lvert d_t \rvert} a_{r}(U_1) a_{s}(U_2) a^{\perp}_{t}(U_3)\, .
\end{equation}
In this expression, the coefficients $\hat{g}_{rs}^{t}$ are the structure constants of the partial permutation algebra $\curlyA_{\infty}$ in the renormalized basis and the $a^{\perp}_{t}$ form the basis of vectors dual to the vectors $a_s$ which means that the inner product between the two bases is a Kronecker delta:
\begin{equation}
    \int dU a_s(U) a_t^{\perp}(U) = \delta_{s,t}\, .
\end{equation}
These vectors $a^{\perp}_{t}$ are explicitly given by linear combinations of the power sum polynomials
\begin{equation}
    a_s^{\perp}(U)=\sum_{k\geq m_{1}(s)} \binom{k}{m_{1}(s)} g_s^{m_{1}(s)-k}  p^*_{\bar{s}\cup 1^k}(U)\, .
\end{equation}
\subsection{The Gromov-Witten Formulation}
The description of the correlation functions of chiral Yang-Mills theory in terms of a partial permutation algebra enables us to provide an explicit string dual by deforming the Gromov-Witten theory. We symbolically define this deformed Gromov-Witten theory as:
\begin{equation}
    \curlyL_{GW} \to \curlyL_{GW} - t\, \tau_1 + \log \Tilde{\Omega}_{u}^{2-2G} \, ,
    \label{StringDualLagrangian}
\end{equation}
where the expression of the exponential distance operator at degree $d$ in terms of Gromov-Witten operators can be written using equation \eqref{SubRule} :
\begin{equation}
    \Tilde{\Omega}_{u} := \sum_{\substack{r\\ \text{minimal}}} u^{\lvert r \rvert} \left(\sum_{\substack{l\leq \lvert d_r \rvert\\ s\, \vdash \, l}} \zeta_{r,s}\, g_s^{\lvert r \rvert-\lvert s \rvert} \tau_{s-1} \right)  \, .
    \label{OmegaFromGW}
\end{equation}
On the one hand, this deformation is implemented at the level of the string theory Lagrangian. Therefore, the factors of the string coupling $g_s$ that appeared in the parameters of the deformations in equation \eqref{ppDeformation} are absorbed by the string expansion. 
On the other hand, the definition of the operator $\Omega$ in terms of 
Gromov-Witten operators \eqref{OmegaFromGW} depends on the string coupling. 
This is a consequence of the fact that the dictionary of observables between 
completed cycles and Gromov-Witten invariants features $g_s$ factors once the genus expansion of the string theory is taken into account \eqref{SubRule}. We stress furthermore that the $\Omega$ operator is defined as a Taylor series in $g_s$ whose dominant term is 
\begin{equation}
    \sum_{\substack{r\\ \text{minimal}}} u^{\lvert r \rvert}  \left( \prod_{i=1}^{\ell(r)} (r_i-1)! \right) \tau_{r-1}  \, .
\end{equation}
Indeed, the coefficients $\zeta_{r,s}$ with $\lvert d_s \rvert < \lvert d_r \rvert$ vanish unless the condition $\lvert s \rvert < \lvert r \rvert$ is satisfied. This can be seen by decomposing the partial permutation orbit $A_r$ in single cycles and showing that the inequality is obeyed by each single cycle.

The need to deform the theory by an operator that includes subdominant corrections in the string coupling suggests that a more appropriate description of the string dual to two-dimensional Yang-Mills theory involves a new compactification of the moduli space of the string theory which is encoded by the exponential distance operator.
Moreover, this deformation imposes that the operators be renormalized by a factor of the inverse operator $\Omega^{-1}$. We denote the renormalized operators by $\hat{\tau}_r:=\Tilde{\Omega}_{u}^{-1} \prod_{i=1}^{\ell(r)} \tau_{r_i}$ and define corresponding vertex operators $\hat{V}_r := g_s^{\ell(r)} \hat{\tau}_r$. The correlation functions of the deformed Gromov-Witten theory are
\begin{equation}
    \left\langle \prod_{i=1}^n \hat{V}_{k_i} \right\rangle  ^{\text{def}} :=\left\langle \left(\prod_{i=1}^n g_s \tau_{k_i}\right) e^{-t\, \tau_1} \Tilde{\Omega}_{u}^{2-2G-n}  \right\rangle\, .
\end{equation}
Restricted to fixed-degrees, they match the correlators of the renormalized conjugacy classes of the deformed Hurwitz theory \eqref{DeformedHurwitz}, 
up to a factor of the string coupling which is absorbed by the genus expansion.
In terms of the Gromov-Witten vertex operators, the correlation functions of chiral Yang-Mills theory read
\begin{equation}
    Z^{+}[M_G,U_i]=\sum_{r_i} \sum_{\substack{s_i\ :\\ \lvert d_{s_i} \rvert\leq  \lvert d_{r_i} \rvert }} \prod_i \Big(a_{r_i}(U) \xi_{r_i,s_i}\Big) \langle \prod_i \hat{V}_{s_i-1} \rangle \, ,
\end{equation}
where we have rescaled the coefficients $\zeta$ to make the string coupling factors associated with each vertex operator manifest: 
\begin{equation}
    \xi_{r,s}=g_s^{\lvert d_r \rvert-\lvert d_s \rvert}\zeta_{r,s}\, .
\end{equation}
The duality map sends a boundary holonomy $U$ in the chiral Yang-Mills theory to the following infinite linear combination of vertex operators in the string theory
\begin{equation}
    \sum_r a_r(U) \sum_{s} \xi_{r,s} \hat{V}_{s-1}\, .
    \label{DualityMap}
\end{equation}
The existence of the map confirms that the string theory we have defined is indeed dual to chiral two-dimensional Yang-Mills theory. 

%

To conclude our description of the string dual of chiral Yang-Mills theory, we stress that on the torus, no deformation of the moduli space of the string theory is required. Therefore, chiral Yang-Mills theory on the torus is exactly dual to the stationary sector of a $T^2$ matter model coupled to topological gravity and deformed by transposition interactions. The dictionary of observables is given by 
\begin{equation}
    U\leftrightarrow \sum_r a_r(U) \sum_{s} \xi_{r,s} \Tilde{\Omega}_u^{-1} V_{s-1}\, .
\end{equation}

\subsection{The Inverse Map}
The duality map \eqref{DualityMap} we have uncovered is a rather complicated one, and it would be interesting to identify new Yang-Mills variables in terms of which which this map simplifies. To that effect, we shall invert the duality map \eqref{DualityMap}. We consider the correlation functions of standard vertex operators on the string theory side 
and derive their expression in terms of Yang-Mills observables. We start by inverting the relation (\ref{SchurToPower}). This yields an expression for the symmetric group characters in terms of the unitary group characters:
\begin{equation}
    \chi_{\lambda}(\alpha)=\int dU s_{\lambda}(U) p_{\alpha}(U^{-1})\, .
\end{equation}
The observables of the deformed Hurwitz theory at fixed degree thus read: 
\begin{equation}
   \langle \prod_{i=1}^k \hat{C}_{\sigma_i} \rangle_{d}^{\text{def}}= \frac{1}{\prod_{i=1}^k  z_{\sigma_i}}\int\left( \prod_{i=1}^k  dU_i\right) \, p_{\sigma_1}(U_1^{-1})\cdots p_{\sigma_k}(U_k^{-1})\cdot Z_{d}^{+}(U_1,\dots,U_k)\, .
    \label{HurwitzAsYM}
\end{equation}
The equation \eqref{HurwitzAsYM} suggests that the basis most natural for comparing operators on both sides is that of the complex-conjugated power sum polynomials $p_{\alpha}^*$: 
\begin{equation}
    \langle \prod_{i=1}^k p^*_{\alpha_i} \rangle_{+} = \int dU_1\cdots dU_k \,  p^*_{\alpha_1}(U)\cdots p^*_{\alpha_k}(U)\, Z^{+}(U_1,\dots,U_k) \, ,
    \label{UpsilonBasis}
\end{equation}
as first argued in \cite{Gross:1993yt}. We note that the insertions must be associated with Young diagrams of the same order to get non-vanishing correlators. In particular, one may replace the full partition function of Yang-Mills in equation \eqref{UpsilonBasis} with the fixed-degree partition function, up to a factor of the chemical potential. We may then write
\begin{equation}
    q^d g_s^{2g(d)-2} \langle \prod_{i=1}^k g_s^{\ell(\sigma_i)} \hat{C}_{\sigma_i} \rangle_{d}^{\text{def}} 
    = \langle \prod_{i=1}^k \frac{p_{\sigma_i}^*}{z_{\sigma_i}} \rangle_{+}\, ,
    \label{BetterHurwitzAsYM}
\end{equation}
where $g(d)$ is the covering surface genus determined by the Riemann-Hurwitz formula for the ramification profiles $\sigma_i$. 
Expressing the correlators of the deformed Gromov-Witten theory in terms of Yang-Mills observables is then a  basis change in the algebra of partial permutations.\footnote{We note that a basis of the family of Yang-Mills observables encoded by the string theory at fixed-degree is given by the correlators $\langle \prod_{i=1}^k p_{\sigma_i}^* \rangle_d$.}. We write
\begin{align*}
    \langle \prod_{i=1}^{k} \hat{\tau}_{k_i} \rangle^{\text{def}}= \langle \Tilde{\Omega}_{g_s u}^{-k} \prod_{i=1}^{k} \frac{\bar{A}_{k_i +1}}{k_i !} \rangle^{\text{def}} \, .
\end{align*}
Writing out the completed cycles explicitly, rearranging the sum and taking out the combinatorial factors arising from the correlators of partial permutations, we obtain
\begin{equation}
    \langle \prod_{i=1}^{k} \hat{\tau}_{k_i} \rangle^{\text{def}}
    =\sum_{s_1,\dots, s_k} \sum_d  g_s^{2g(d)-2} q^d  \left(\prod_{i=1}^{k} \binom{m_1(\sigma_i)}{m_1(s_i)}\, \frac{\rho_{k_i+1,s_i}}{k_i !}\right)\langle \prod_{i=1}^{k}  \hat{C}_{\sigma_i} \rangle_{d}^{\text{def}} \, .
\end{equation}
Substituting in equation \eqref{BetterHurwitzAsYM}, we find
\begin{equation}
   \langle \prod_{i=1}^{k} \hat{\tau}_{k_i} \rangle^{\text{def}} = \sum_d \sum_{s_1,\dots, s_k} \left(\prod_{i=1}^{k}  \frac{g_s^{-\ell(\sigma_i)}}{k_i! z_{\sigma_i}} \binom{m_1(\sigma_i)}{m_1(s_i)} \rho_{k_i+1,s_i}\right) \langle \prod_{i=1}^k p_{\sigma_i}^* \rangle_{+}
   \label{YMSector}\, .
\end{equation}
This equation is the inverse of \eqref{DualityMap}: it describes the Yang-Mills observables naturally computed by standard string theory correlators. These observables are a completed version of the power sum polynomials.
It is interesting to  manipulate the expression a little further to uncover the role of the trivial sheets:
\begin{equation}
   \langle \prod_{i=1}^{k} \hat{\tau}_{k_i} \rangle^{\text{def}}  = \left\langle \sum_{s_1,\dots, s_k} \left(\prod_{i=1}^{k} \frac{\rho_{k_i+1,s_i}}{k_i!} \frac{g_s^{-\ell(s_i)}}{z_{s_i}}\, p^*_{s_i}\right) \sum_d \left(\prod_{i=1}^k \frac{g_s^{-(d-\lvert d_{s_i} \rvert)}}{(d-\lvert d_{s_i} \rvert)!}\right) (p_{1}^*)^{d-\lvert d_{s_i} \rvert} \right\rangle_{+}\, .
\end{equation}
They give rise both to operator insertions $p_1^\ast$ as well as combinatorial factors. 
%

\subsection{Remarks on Wilson Loops}
\label{WilsonLoops}
In the previous subsections, we have studied partition functions of the chiral Yang-Mills theory. We have shown that in the absence of boundaries, the partition functions are reproduced by a deformed Gromov-Witten theory of the target space. Furthermore, we have argued that boundaries with prescribed holonomies may be thought of as operator insertions, and that their correlation functions match the corresponding Gromov-Witten correlators. In this subsection, we study correlators of Wilson lines in chiral Yang-Mills theory. As an opening remark, we observe that the expectation value of a Wilson loop depends on the area it encloses. Therefore, the area deformation of the Gromov-Witten theory of the target space must be patch-dependent: we view the parameter $t$ as an operator which associates its area to the patch under consideration. 

Wilson lines on a manifold of genus $G$ and area $t$ are computed by gluing manifolds of genus $G$ and zero, with respective area $t_1$ and $t_2=t-t_1$, along a boundary with an extra character insertion:
\begin{equation}
    \langle s_{\mu} \rangle_G = \int dU Z^{+}_{G,t_1} (U) s_{\mu}(U) Z^{+}_{0,t_2} (U^{-1})\, .
\end{equation}
As we explained in section \ref{StringDual}, one should think of the functions
$Z^{+}_{G,t_1} (U)$ and $Z^{+}_{0,t_2}(U^{-1})$ as Gromov-Witten invariants relative to the boundary with prescribed holonomy $U$. 
If we rewrite such Wilson line expectation values in terms of symmetric group quantities, we find:
\begin{multline}
    \langle s_{\mu} \rangle_G =  \frac{d_{\mu}}{d !}\, \sum_{d_1,d_2} g_s^{d_1 (2G-1)} g_s^{-d_2} e^{-\frac{t_1 d_1+t_2  d_2}{2}}  \sum_{\substack{\lambda\, \vdash d_1\\ \rho\, \vdash d_2}} \left(\frac{d_{\lambda}}{d_1!}\right)^{2-2G} \left(\frac{d_{\rho}}{d_2 !}\right)^2 \Omega_{g_s}(\lambda)^{1-2G}
    \\
    \Omega_{g_s}(\rho)\  e^{-g_s (t_1 f_{2}(\lambda)+t_2 f_{2}(\rho))}
    \sum_{\substack{\alpha\, \vdash d_1\\ \beta\, \vdash d \ }} z_{\alpha\cup \beta} f_{\alpha}(\lambda) f_{\beta}(\mu)  f_{\alpha\cup \beta}(\rho) \delta_{d_1+d,d_2}\, ,
\end{multline}
where $\mu$ is a partition of $d$ and $\alpha\cup\beta$ denotes the concatenation of the conjugacy classes $\alpha$ and $\beta$. This expression may be simplified to render the gluing method more manifest on the symmetric group side of the equation: 
\begin{equation}
    \langle s_{\mu} \rangle_G = \left(\frac{e^{-\frac{t_2}{2}}}{g_s}\right)^d \, \sum_{\beta\, \vdash d} \chi_{\mu}(\beta) \Bigg( \sum_{d_1} g_s^{d_1(2G-2)}  e^{-\frac{t d_1}{2}} \sum_{\alpha\, \vdash d_1} \frac{z_{\alpha\cup \beta}}{z_{\beta}}  \langle C_{\alpha} \rangle_{d_1, G}^{t_1,u}  \langle C_{\alpha \cup \beta} \rangle_{d+d_1, 0}^{t_2,u}\Bigg) \, .
    \label{WilsonLineGluing}
\end{equation}
%
Equation (\ref{WilsonLineGluing}) provides a generic gluing prescription for how to incorporate a Wilson line in the Hurwitz formulation of the Yang-Mills theory. This prescription involves a sum over a basis of intermediate states corresponding to conjugacy classes. 
A direct string theoretic interpretation of the gluing procedure would be interesting. In particular, it would be neat to interpret the Wilson lines as an extended object in the dual string theory.

\section{Conclusions}
\label{Conclusions}
It was well-known that two-dimensional Yang-Mills theory allows for a large $N$ expansion that resembles a string theory. A sum over covering maps reproduces the expansion and the description of the covering maps of degree $d$ is closely related to the description of $S_d$ principal bundles that are part of the configuration space of
a two-dimensional topological quantum field theory. There were several open questions in the gauge/string duality. We have attempted to address some of them using new tools.
We introduced a Frobenius algebra of partial permutations of infinite order which codes $S_d$ bundles that describe the covering for every possible degree $d$ of the map. %
The second tool we used is the Hurwitz/Gromov-Witten correspondence. We stressed that this correspondence is between a grand canonical Hurwitz theory (or infinite order partial permutation algebra) and a bona fide string theory described by two-dimensional topological matter coupled to two-dimensional gravity. The correspondence thus allows us to rewrite two-dimensional Yang-Mills observables in terms of a perturbatively well-defined  string theory
incorporating an integral over the moduli space of Riemann surfaces. These two tools clarify considerably the structure of the string theory dual to two-dimensional Yang-Mills theory. 

Taking the large $N$ limit is a necessary procedure to establish a gauge/string duality and it renders the correspondence perturbative. This may be expected as we are more familiar with a perturbative definition of world-sheet string theories defined in a genus expansion weighted by the string coupling $g_s=1/N$. Of course, we are allowed to think of the two-dimensional Yang-Mills theory as a non-perturbative completion of the dual perturbative string theory. We think this is a potentially powerful perspective worth fleshing out further. 

The perturbative string theory we defined is a deformation of a Gromov-Witten theory. We deform on the one hand by an operator that has also been interpreted as a compactification of Hurwitz spaces. On the other hand, we deform by a transposition operator proportional to the area of the target space.
The latter deformation is standard in introducing the simplest interaction in a string theory e.g. in matrix string theory \cite{Dijkgraaf:1997vv} or  in proposals for the dual of an $AdS_3$ string theory \cite{Eberhardt:2021vsx}. 

Indeed, several aspects of our two-dimensional gauge/string correspondence are reminiscent of $AdS_3/CFT_2$. There is for instance the grand canonical nature of the field theory dual to the string theory which is also present in the symmetric orbifold theory dual to Neveu-Schwarz-Neveu-Schwarz $AdS_3$  string theory \cite{Kim:2015gak,Eberhardt:2020bgq}.
Here, we saw that the grand canonical aspect is also present in the intermediate extended Hurwitz theory that arises in the large $N$ limit of two-dimensional Yang-Mills theory.

There are multiple avenues of further research. One would be to gain a better understanding of the compactification of Hurwitz space encoded in the $\Omega$ deformation. It would also be interesting to formulate an action principle for the deformed Gromov-Witten theory that we have defined. Another avenue is the characterization of the Wilson line operators in the dual string theory. In the other direction of the duality, we can ask what the counterpart is of operators like the gravitational descendants of the identity operator on the gauge theory side. 
These are but a few questions that have become much more concrete through our detailed description of the archetypal two-dimensional gauge/string theory duality. 

\appendix
\section{Useful Identities}
\label{Useful Identities}
In this appendix, we collect useful identities on symmetric polynomials and their inner products. Firstly, we recall that the character of a unitary matrix $s_{\lambda}(U)$ is a Schur polynomial evaluated in the eigenvalues of the representation $\lambda$ of the matrix $U$. Schur polynomials are related to power sum polynomials through
\begin{equation}
    p_{\alpha}(U)=\sum_{\lambda \in Y^{d}} s_{\lambda}(U) \chi_{\lambda} (\alpha)\, ,
\end{equation}
where $d=\sum_i \alpha_i$ and the $p_{\alpha}(U)$ are the power sum polynomials evaluated at eigenvalues of  the representation $\lambda$ of the matrix $U$. This relation can be inverted:
\begin{equation}
        s_{\lambda}(U) =\sum_{\alpha \in Y^{d}} \frac{1}{z_{\alpha}} p_{\alpha}(U) \chi_{\lambda} (\alpha)\, ,
\end{equation}
where $z_{\alpha}$ is the cardinal of the centralizer of the permutation $\alpha$ and $d=\sum_i \lambda_i$. The orthogonality relations satisfied by the characters of a group may be exploited to derive the following orthogonality relation for the power sum polynomials:
\begin{align}
    \int dU p_{\alpha}(U) p_{\beta}(U^{-1}) &=\sum_{\lambda,\rho} \chi_{\lambda}(\beta)  \chi_{\rho}(\alpha) \int dU s_{\lambda}(U) s_{\rho}(U^{-1})\\
    &=\sum_{\lambda} \chi_{\lambda}(\alpha)  \chi_{\lambda}(\beta) = z_{\alpha} \delta_{\alpha, \beta}\, .
    \label{OrthoRelations}
\end{align}

\section{Determining the Dictionary}
\label{Solving the Dictionary}
In order to determine the coefficients \eqref{DictionarySolution}, one can fix some maximal degree and solve the linear system of equations up to that degree by Gaussian elimination. This yields the ansatz \eqref{DictionarySolution} which may then be proved by induction. Here, we shall simply check that the ansatz \eqref{DictionarySolution} does indeed solve equation \eqref{DictionaryCoeffEqs}. Uniqueness of the solution follows by restricting to finite values for the degree. Inside correlation functions of degree $d$, 
$\sum_{r} a_r(U) A_{r}$ behaves as 
\begin{equation}
    \sum_{\rho} \left( \sum_{k=0}^{m_{1}(\rho)} a_{\bar{r}\cup 1^k}(U) g_s^{\ell(\bar{r})+k} \binom{m_{1}(\rho)}{k} \right) C_{\rho}\, ,
    \label{DegreeBehaviour}
\end{equation}
where $m_{1}(\rho)=d-\lvert d_{\bar{r}}\rvert$. Substituting \eqref{DictionarySolution} in \eqref{DegreeBehaviour}, we find that the coefficients of each $C_{\rho}$ are given by 
\begin{equation}
    p_{\bar{r}}(U) \sum_{k=0}^{m_{1}(\rho)} \sum_{l=0}^{k} (-1)^{k-l}  g_{s}^{l+\ell(\bar{r})} \binom{k}{l} \binom{m_{1}(\rho)}{k}  p_{1^l}(U)\, .
\end{equation}
We proceed to reorder this sum. We fix a non-negative integer $i\leq m_{1}(\rho)$ and compute the coefficient of $p_{1^i}(U)$ in the above sum. It is given by
\begin{equation}
    p_{\bar{r}}(U) g_{s}^{i+\ell(\bar{r})} \sum_{k=i}^{m_{1}(\rho)} (-1)^{k-i} \binom{k}{i} \binom{m_{1}(\rho)}{k}\, .
\end{equation}
We then shift the index $k$ of the sum such that it starts at $0$. We thus define $j:=k-i$ and write the coefficient of $p_{1^i}(U)$ as 
\begin{equation}
    p_{\bar{r}}(U) g_{s}^{i+\ell(\bar{r})} \sum_{j=0}^{m_{1}(\rho)-i} (-1)^{j} \binom{i+j}{i} \binom{m_{1}(\rho)}{i+j} \, .
\end{equation}
We exploit an identity for binomial coefficients to rewrite the coefficient of $p_{1^i}(U)$ as
\begin{equation}
    p_{\bar{r}}(U) g_{s}^{i+\ell(\bar{r})} \binom{m_{1}(\rho)}{i} \sum_{j=0}^{m_{1}(\rho)-i} (-1)^{j} \binom{m_{1}(\rho)-i}{j}\, .
\end{equation}
The alternate sum of binomial coefficients $\sum_k (-1)^k \binom{n}{k}$ is zero for all $n\geq 1$. Therefore, the only value of $i$ for which the coefficient of $p_{1^i}(U)$ does not vanish is $i=m_{1}(\rho)$. We conclude that
\begin{equation}
    \sum_{k=0}^{m_{1}(\rho)} a_{\bar{r}\cup 1^k}(U) g_s^{\ell(\bar{r})+k} \binom{m_{1}(\rho)}{k} = g_{s}^{\ell(\rho)} p_{\rho}(U)\, ,
\end{equation}
which is what we had set out to prove.

\section{Phase Transitions}
\label{PhaseTransitions}
Two-dimensional Yang-Mills theory admits a phase transition on the sphere \cite{Douglas:1993iia}. Indeed, its free energy trivializes for small coupling, namely $g_{YM}^2 A < \pi^2$. This transition was studied in \cite{Gross:1994mr} from the perspective of a weak-coupling expansion and was understood in that paper to originate in instanton corrections. More precisely, only representations of the unitary group with a small number of boxes contribute to the free energy below the critical value of the Yang-Mills coupling.
One thus expects the string theory to break down in this limit and become a theory of point particles. It would be interesting to understand this phase transition from the dual string point of view. A first step in this direction was taken in \cite{Taylor:1994zm,Crescimanno:1994eg}, where phase transitions in various simplifications of the string dual of two-dimensional Yang-Mills theory were analyzed through the study of the radius of convergence of the free energy. The free energy of the chiral Yang-Mills string was shown to converge for small areas and large areas, but not in the intermediate regime. This suggests the existence of three distinct phases for the dual string. In this section, we provide a broader analysis of phase transitions in the generalized chiral Yang-Mills string, first analytically, then numerically.
\subsection{Analytical Phase Transitions}
The free energy of the chiral string theory is defined as the logarithm of the disconnected partition function of the theory. It may be identified with a partially evaluated generating function of disconnected double Hurwitz numbers. Specifically, the parameters associated with the non-trivial insertions are evaluated to $1$ and the instanton weight is related to the area. 
We slightly extend this framework by defining the deformed string theory with Lagrangian \eqref{StringDualLagrangian} for all values of the parameters $t$ and $u$. 
In this perspective, the chiral Yang-Mills theory is a special line in a three-dimensional space of parameters indexed by $(q\leq 1,t,u)$. Its partition function reads
\begin{equation}
    Z(q,g_s,t,u)= \sum_g g_{s}^{2g-2} \sum_{d} q^d \sum_{r} \frac{(-1)^r t^r}{r!} \sum_{\alpha,\beta\vdash d} u^{\lvert \alpha \rvert} u^{\lvert \beta \rvert} H_{g,d}^{\bullet} (C_{\alpha},C_{\beta},C_2^r)\, ,
\end{equation}
and it computes the generating function for double Hurwitz numbers with a single parameter $u$ for the two non-trivial insertions. As a result, the logarithm $F$ of the partition function generates connected Hurwitz numbers \cite{Novak:2023bft, CavalieriMiles}. We find
\begin{equation}
    F(q,g_s,t,u)= \sum_g g_{s}^{2g-2} \sum_{d} q^d \sum_{r} \frac{(-1)^r t^r}{r!}\, H_{g,d}^{\circ}(\Omega_{u}^2,C_2^r)\, ,
\end{equation}
where $H_{g,d}^{\circ}(\cdot)$ denotes connected Hurwitz numbers with specified
insertions for degree $d$ and covering surface genus $g$. An interesting property of 
this free energy is that it has a  built-in redundancy. By expanding the $\Omega$ points and leveraging the Riemann-Hurwitz constraint, one can show the invariance under scaling
\begin{equation}
    F(\alpha^{-1} g_s, \alpha^{2G-2} q, \alpha t, \alpha u) =  F(q,g_s,t,u)\, ,
    \label{Symmetry}
\end{equation}
where $\alpha$ is non-zero. 
In other words, the free energy depends only on three parameters $\tilde{g_s}=g_s u$, $\tilde{q}=q u^{2-2G}$ and $\tilde{t}:=t/u$, provided $u$ is non-zero.\footnote{We note that, if $u$ is zero, one can follow the same procedure to remove another parameter.} We write $F(g_s,q,t,u)=F(\tilde{g_s}, \tilde{q},\tilde{t},1)$. 
This symmetry is a generic feature of generating functions of Hurwitz numbers, and it can be traced back to the existence of the Riemann-Hurwitz constraint. It simplifies considerably the study of phase transitions in the theory. In particular, the phase diagrams of the theory must preserve this symmetry. 

We now study the dominant contribution to the genus expansion of the free energy, namely the genus $0$ term. It is this term which determines whether there is a phase transition \cite{Douglas:1993iia}. We define
\begin{equation}
    F_0(q,t,u):=\lim_{g_s\to 0} g_s^2 F(q,g_s,t,u)\, .
\end{equation}
Our strategy, following \cite{Taylor:1994zm}, consists in determining the radius of convergence of the free energy and interpreting the results as describing different 
phases of the theory. We start with a discussion of the phase transitions in the $(q,t)$ and $(q,u)$ planes. In the absence of $\Omega$ insertions, namely for a Hurwitz theory deformed by single branch points only, one has
\begin{equation}
     F_0(q,t,0)=\sum_{d} q^d \sum_r \frac{(-1)^r t^r}{r!} H_{0,d}^{\circ}(C_2^r)\, .
\end{equation}
The Riemann-Hurwitz formula imposes that the number of single branch points be $r=2d-2$.
The free energy thus reads
\begin{equation}
     F_0(q,t,0)=\sum_{d} q^d \frac{t^{2d-2}}{(2d-2)!} H_{0,d}^{\circ}(C_2^{2d-2})\, .
\end{equation}
The simple, genus $0$ Hurwitz number on the right-hand-side evaluates to $\frac{(2d-2)!}{d!} d^{d-3}$ \cite{HurwitzPaper, Lando}. Using the Stirling formula to evaluate the factorial asymptotically, one obtains
\begin{equation}
     F_0(q,t,0)\sim \frac{1}{\sqrt{2\pi}} \frac{1}{t^2} \sum_{d} \frac{(q e t^2)^d}{d^{7/2}}\, .
\end{equation}
The free energy for a grand canonical Hurwitz theory deformed by single branch points thus converges for 
\begin{equation}
    t^2 < \frac{1}{e\cdot q}\, .
\end{equation}
This result generalizes slightly that of \cite{Taylor:1994zm}. The transition occurs when the weight given to each single branch point exceeds a certain threshold. If $t$ is interpreted as the area that a single branch point may occupy, i.e. the entropy of a single branch point, then one finds that it is the entropy of single branch points which drives the phase transition \cite{Taylor:1994zm}. 

Consider now the $\Omega$-deformed Gromov-Witten theory. The dominant contribution free energy reads
\begin{equation}
    F_0(q,0,u)=\sum_d q^d \sum_{\alpha,\beta} u^{\lvert \alpha \rvert + \lvert \beta \rvert}H_{0,d}^{\circ}(C_{\alpha},C_{\beta}) \, .
\end{equation}
Applying again the Riemann-Hurwitz formula, we find that the only conjugacy class that contributes are the cycles of maximal length $C_d$. Furthermore, the connected Hurwitz number $H_{0,d}^{\circ}(C_d,C_d)$ matches the disconnected one, because the action of $C_d$ on the set $\{1,\dots,d\}$ is transitive. This Hurwitz number is
\begin{equation}
    H_{0,d}^{\circ}(C_d,C_d)=\frac{\lvert C_d \rvert}{d!}=\frac{1}{d} \, .
\end{equation}
The free energy thus reads
\begin{align}
     F_0(q,0,u) &=\frac{1}{u^2} \sum_d \frac{(q u^{2})^d}{d}\\
     &= -\frac{1}{u^2} \log(1-qu^2) \, .
     \label{TopologicalFreeEnergy}
\end{align}
The free energy thus converges for $u^2<1/q$, namely when the weight associated with the $\Omega$ deformation is not too large compared to the Yang-Mills coupling. For the value of the deformation parameter which reproduces the chiral Yang-Mills theory when single branch points are added, i.e. for $u=1$, the free energy does not exhibit a divergence, 
in accordance with the results of \cite{Taylor:1994zm}.

The case of the full chiral string is harder to deal with analytically. One has
\begin{align}
    F_0(q,t,u) 
    &=\sum_{d} q^d \sum_{\alpha,\beta\, \vdash d} (-1)^{\ell(\alpha)+\ell(\beta)-2}  \frac{t^{\ell(\alpha)+\ell(\beta)-2}  u^{\lvert \alpha \rvert+\lvert \beta\rvert} }{(\ell(\alpha)+\ell(\beta)-2)!}H_{0,d}^{\circ}(C_{\alpha},C_{\beta},C_2^{\ell(\alpha)+\ell(\beta)-2}) \, .
    \label{GeneralizedFreeEnergy}
    %
\end{align}
The free energy is thus a series in three parameters. For each value of $d$, the summand is given by a homogeneous polynomial of degree $2d-2$ in the variables $t$ and $u$, weighted by $q^d$. The coefficients of the pure monomials $t^{2d-2}$ and $s^{2d-2}$ are respectively the Hurwitz number $H_{0,d}^{\circ}(C_2^{\ell(\alpha)+\ell(\beta)-2})$ and $H_{0,d}^{\circ}(C_{\alpha},C_{\beta})$, which we have computed above. The work \cite{Taylor:1994zm} provided a numerical analysis which suggests that the chiral Yang-Mills string, for which $q=e^{-t/2}$ and $u=1$, admits three distinct phases. 
Here, we state the precise mathematical problem that must be solved in order to draw the precise phase diagram of the generalized chiral Yang-Mills string. 
The problem is to determine the radius of convergence of the generating function of connected double Hurwitz numbers which was studied in \cite{Okounkov:2000fna, OP1}. 

\subsection{Numerical Phase Transitions}
At present, computing double Hurwitz numbers remains largely outside the scope of analytical methods. Numerically, disconnected double Hurwitz numbers can be computed for small values of the degree using the definition of Hurwitz numbers in terms of conjugacy classes. However, computing the free energy 
\eqref{GeneralizedFreeEnergy} of the deformed extended Hurwitz theory remains hard. In the case of Yang-Mills, one expects that the qualitative behavior of the phase diagram of the theory is also encoded in the behavior of its one-point functions \cite{Gross:1994mr}. 
Therefore, we shall compute certain one-point functions of the deformed extended Hurwitz theory. We will then attempt to draw lessons concerning the phase diagram of the free energy of the chiral Yang-Mills theory. The one-point function of the chiral Yang-Mills theory on the sphere, or equivalently the free energy on the disk, reads
\begin{equation}
    F_0(\hat{A}_{\bar{r}};q,t,u) = \sum_d q^d \sum_{\alpha \vdash d} \frac{(-1)^{\ell(\rho) + \ell(\alpha)}}{(\ell(\rho) + \ell(\alpha) - 2)!} t^{\ell(\rho) + \ell(\alpha) - 2} u^{|\alpha|} H_{0, d}^{\circ}(C_{\rho}, C_{\alpha}, C_{2}^{\ell(\rho) + \ell(\alpha) - 2})\, .
\end{equation}
The operator $\hat{A}_{\bar{r}}$ may be thought of as a complicated insertion in the chiral Yang-Mills theory. Note that by trading the sphere for a disk, we have traded an infinite sum over all conjugacy classes for a single class insertion. Nevertheless, the above expression cannot easily be computed in all generality, since it still features a sum over double Hurwitz numbers. We thus restrict our attention to the insertion of the renormalized identity and transposition classes. We write
\begin{align}
    F_0(\hat{A}_{\varnothing};q,t,u) &= \sum_d q^d \sum_{\alpha \vdash d} \frac{(-1)^{\ell(\alpha)+d}}{(\ell(\alpha)+d-2)!} t^{\ell(\alpha)+d-2} u^{|\alpha|} H_{0, d}^{\circ}(C_{\alpha}, C_{2}^{\ell(\alpha)+d-2})\\
    F_0(\hat{A}_{2};q,t,u) &= -\sum_d q^d \sum_{\alpha \vdash d} \frac{(-1)^{\ell(\alpha)+d}}{(\ell(\alpha)+d-3)!} t^{\ell(\alpha)+d-3} u^{|\alpha|} H_{0,d}^{\circ}(C_{\alpha}, C_{2}^{\ell(\alpha)+d-2})\, .
\end{align}
Only connected, single Hurwitz numbers at genus $0$ appear in these expressions and they admit the following exact expression \cite{HurwitzPaper, Lando}:
\begin{equation}
    H_{0, d}^{\circ}(C_{\alpha}, C_{2}^{d + \ell(\alpha) - 2}) = \frac{(d + \ell(\alpha) - 2)!}{|\text{Aut}(\alpha)|} \prod_{i = 1}^{\ell(\alpha)} \frac{\alpha_i^{\alpha_i}}{\alpha_i!} d^{\ell(\alpha)-3}.
\end{equation}
Furthermore, given that we are interested in the case where $u\neq 0$, we may exploit the symmetry \eqref{Symmetry} to reduce the number of parameters that each one-point function depends on. We write 
\begin{align}
    F_0(\hat{A}_{\varnothing};q,t,u) &= u^{-2} \sum_d (qu^2)^d \sum_{\alpha \vdash d} (-1)^{\ell(\alpha)+d}  (t/u)^{\ell(\alpha)+d-2}\cdot \frac{1}{|\text{Aut}(\alpha)|} \prod_{i = 1}^{\ell(\alpha)}  \frac{\alpha_i^{\alpha_i}}{\alpha_i!} d^{\ell(\alpha)-3}.\\
    F_0(\hat{A}_{2};q,t,u) &= -u^{-3} \sum_d (q u^2)^d \sum_{\alpha \vdash d} (-1)^{\ell(\alpha)+d} (t/u)^{\ell(\alpha)+d-3} \cdot \frac{d + \ell(\alpha) - 2}{|\text{Aut}(\alpha)|} \prod_{i = 1}^{\ell(\alpha)} \frac{\alpha_i^{\alpha_i}}{\alpha_i!} d^{\ell(\alpha)-3}\, .
\end{align}
The overall factors of $u$ are irrelevant to the study of the radius of convergence of the partition functions since $u\neq 0$. Therefore, we focus on the sums and define the new parameters $q'=qu^2$ and $t'=t/u$. The partition functions start to oscillate with increasing magnitude when $q'$ and $t'$ become large enough, indicating that the one-point function becomes divergent in this region. We plot here the regions in coordinate space for which the partition functions seem to converge.
\begin{figure}[H]
    \centering
    \includegraphics[width=5.5cm]{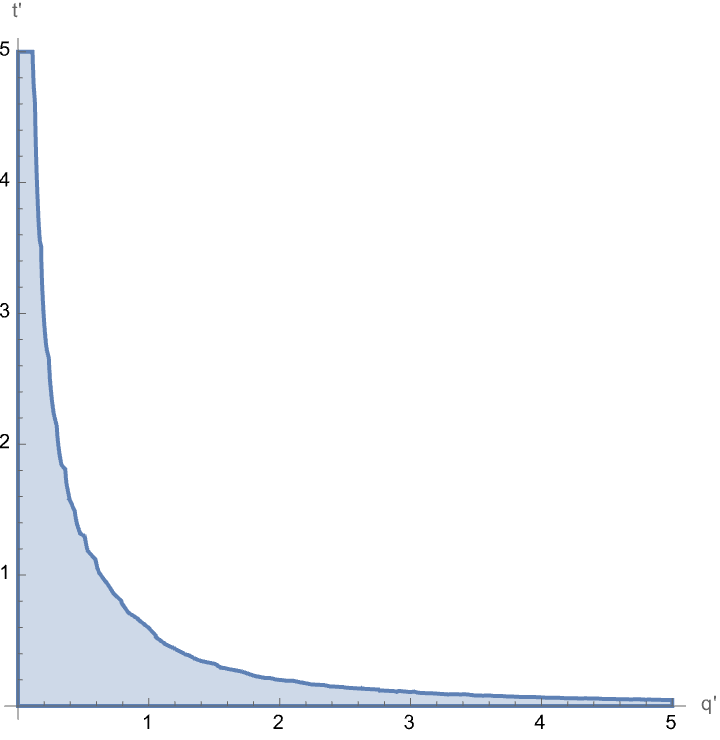}
    \includegraphics[width=5.5cm]{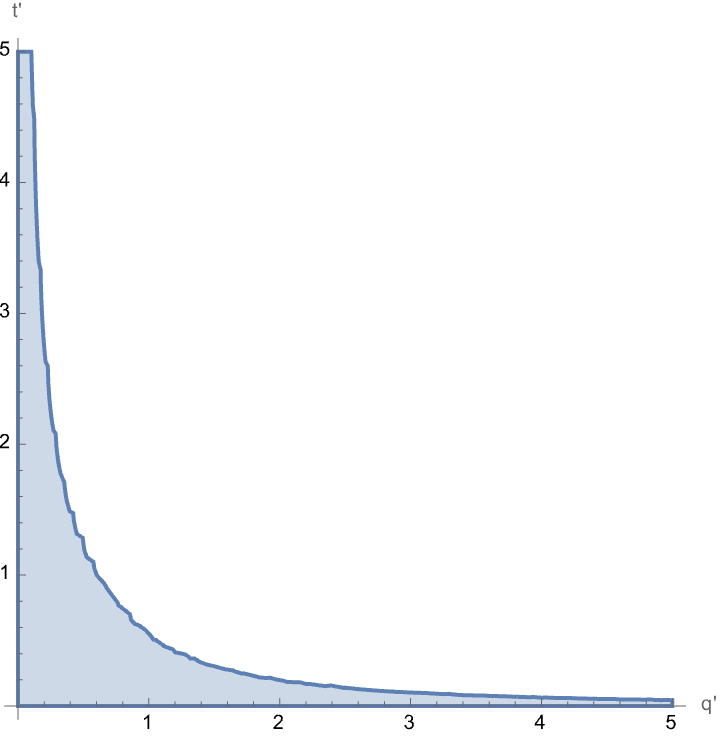}
    \caption{The domains of convergence of the renormalized identity one-point function (on the left) and of the renormalized transposition one-point function (on the right).}
    \label{DomainsOfConvergence}
\end{figure}
The line in parameter space corresponding to chiral Yang-Mills theory is $(q'=e^{-t'/2},t')$. We plot the correlation function of the renormalized identity class as
a function of the area $t$ of the target space sphere on the Yang-Mills line of parameter space.
\begin{figure}[H]
    \centering
    \includegraphics[width=7cm]{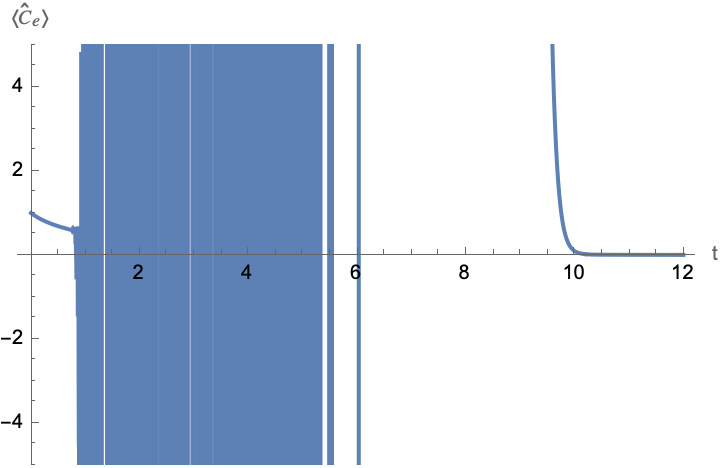}
    \caption{The one-point function for a renormalized identity class insertion as a function of $t$.}
    \label{ChiralYMPlot}
\end{figure}
The partition function appears to converge for small and large values of the area, and exhibits rapid, high amplitude oscillations in an intermediate range, indicating a divergence. This strongly suggests that the chiral Yang-Mills theory  exhibits three phases, in accordance with \cite{Taylor:1994zm}. The phase transitions occur around $t\sim 0.8$ and $t\sim 10$, respectively.\footnote{ Playing with the degree to which the computation is carried out suggests that these are  rough values, and only the qualitative behavior of the one-point functions should be trusted.}
Figure \ref{DomainsOfConvergence} therefore constitutes a qualitative phase diagram for the generalization of the chiral Yang-Mills theory that we have defined.

\bibliographystyle{JHEP}

\end{document}